\documentclass[preprint,12pt]{elsarticle}
\usepackage{amsmath}
\usepackage{algorithmic}
\usepackage{algorithm}
\usepackage{array}
\usepackage{subfig}
\usepackage{caption}
\usepackage{stfloats}
\usepackage{url}
\usepackage{verbatim}
\usepackage{graphicx}
\usepackage{amssymb,amsthm,enumerate}
\usepackage{threeparttable}
\usepackage{mathrsfs}
\usepackage{color}
\usepackage{epstopdf}
\usepackage{indentfirst}
\usepackage[latin1]{inputenc}
\usepackage{rotating}
\usepackage{colortbl}
\usepackage{bm}
\usepackage[T1]{fontenc} 
\usepackage{mathtools}
\usepackage{booktabs}
\usepackage{tikz}
\usepackage{verbatim}
\usepackage{graphicx}
\usepackage{amssymb,amsthm,enumerate}
\usepackage{threeparttable}
\usepackage{mathrsfs}
\usepackage{color}
\usepackage{epstopdf}
\usepackage{indentfirst}
\usepackage[latin1]{inputenc}
\usepackage{rotating}
\usepackage{colortbl}
\usepackage{bm}
\usepackage[T1]{fontenc} 
\usepackage{mathtools}
\usepackage{booktabs}

\newtheorem{theorem}{Theorem}
\newtheorem{corollary}{Corollary}
\newtheorem{definition}{Definition}

\newtheorem{assumption}{Assumption}

\newtheorem{remark}{Remark}

\journal{Journal of \LaTeX\ Templates}

\begin{document}

\begin{frontmatter}

	\title{ Dynamics-Based Algorithm-Level Privacy Preservation for Push-Sum Average Consensus }

	\author[]{Huqiang Cheng$^{a}$}
	\ead{huqiangcheng@126.com}

	\author[]{Mengying Xie$^{a,*}$}
	\ead{xiemyscut@163.com}

	\author{Xiaowei Yang$^{b}$}
	\ead{xwyang@scut.edu.cn}

	\author[]{Qingguo L{\"u}$^{a}$}
	\ead{qglv@cqu.edu.cn}

	\author[]{Huaqing~Li$^{c}$}
	\ead{huaqingli@swu.edu.cn}

	\address[a]{Key Laboratory of Dependable Services Computing in Cyber Physical Society-Ministry of Education, College of Computer Science, Chongqing University, Chongqing, 401331, China}
	\address[b]{School of Software Engineering, South China University of Technology, Guangzhou, 510006, China}
	\address[c]{Chongqing Key Laboratory of Nonlinear Circuits and Intelligent Information Processing, College of Electronic and Information Engineering, Southwest University, Chongqing, 400715, China}

	\cortext[mycorrespondingauthor]{Corresponding author.}

	\begin{abstract}
In the intricate dance of multi-agent systems, achieving average consensus is not just vital--it is the backbone of their functionality. In conventional average consensus algorithms, all agents reach an agreement by individual calculations and sharing information with their respective neighbors. Nevertheless, the information interactions that occur in the communication network may make sensitive information be revealed. In this paper, we develop a new privacy-preserving average consensus method on unbalanced directed networks. Specifically, we ensure privacy preservation by carefully embedding randomness in mixing weights to confuse communications and introducing an extra auxiliary parameter to mask the state-updated rule in initial several iterations. In parallel, we exploit the intrinsic robustness of consensus dynamics to guarantee that the average consensus is precisely achieved. Theoretical results demonstrate that the designed algorithms can converge linearly to the exact average consensus value and can guarantee privacy preservation of agents against both honest-but-curious and eavesdropping attacks. The designed algorithms are fundamentally different compared to differential privacy based algorithms that enable privacy preservation via sacrificing consensus performance. Finally, numerical experiments validate the correctness of the theoretical findings.
	\end{abstract}

	\begin{keyword}
		{Average consensus, privacy preservation, tailored weights, unbalanced directed networks}
	\end{keyword}

\end{frontmatter}


\section{Introduction}
Recently, multi-agent systems have been rapidly expanding in industrial development. A key characteristic of these systems is that the various agents collaborate to achieve a consensus state. The average consensus state is a crucial evaluation metric for multi-agent systems, leading to the development of numerous average consensus algorithms. Considering a network with $N$ agents, the goal of such algorithms is to make the states of all agents converge asymptotically to the average of their initial values. Due to the inherent decentralized characteristics, average consensus approaches are widely used in many areas, such as collaborative filtering \cite{Bae2021}, decision-making system \cite{Yang2012}, social network \cite{Zhang2020}, UAV formation \cite{Souza2022}, online learning \cite{Wu2024}, etc.

In order to make the states of all agents reach the average of initial values, most of average consensus approaches \cite{Kempe2003,Xia2023,Benezit2010,Hadjicostis2018,Liu2024a,Guo2024} always demand that the agents share their correct states with each other. This may result in privacy information being revealed, and it is highly inadvisable from the perspective of privacy protection. Privacy concerns in multi-agent systems are of great significance in daily life. A simple example is a group of individuals engaging in a discussion regarding a specific topic and reaching a common view while maintaining the confidentiality of each individual view \cite{Tsitsiklis1984}. A further common example is in power systems where several generators need to agree on costs as well as ensuring the confidentiality of their respective generation information \cite{Zhang2011}. As the frequency of privacy breaches continues to rise, it has become increasingly urgent to safeguard the privacy of every individual in multi-agent systems.

\subsection{Related Works}
Several approaches have been available to tackle the growing privacy concerns in average consensus literature. One of the mostly widespread non-encryption privacy-preserving techniques is differential privacy \cite{Dwork2006}, which essentially injects the uncorrelated noise to the transmitted state information. This strategy has already been applied in \cite{Huang2015,Nozari2017,Huang2012,Gao2021,Ye2020}. However, such approaches cannot achieve exact average consensus owing to its inherent privacy-accuracy compromise. This makes differentially private approaches unpalatable for sensor networks and cyber-physical systems with high requirements for consensus accuracy. To ensure computational accuracy, several improvement efforts were developed in \cite{Kefayati2007,Mo2017,Manitara2013}, which focus on the strategic addition of correlated noise to the transmitted information, as opposed to the uncorrelated noise typically utilized in differential privacy. Another stand of interest is observability-based privacy-preserving approaches \cite{Pequito2014,Ridgley2019,Alaeddini2017}, where the privacy is guaranteed by minimizing the observation information of a certain agent. However, both the correlated-noise based and the observability-based approaches are vulnerable to external eavesdroppers who have the ability to wiretap all communication channels.

Note that the above mentioned approaches are only valid for undirected and balanced networks. In real-world scenarios, communication among agents is usually directed and unbalanced. For example, broadcasting at different power levels, the communication activity corresponds to a directed and unbalanced networks. To preserve privacy of nodes interacting on an unbalanced directed network, the authors in \cite{Kishida2018,Hadjicostis2020,Fang2021,Ruan2019} presented a series of encryption-based approaches by utilizing the homomorphic encryption techniques. However, this type of approaches requires substantial computational and communication overhead, which is unfriendly to resource-limited systems. Recently, state-decomposition based approaches \cite{Wang2019,Chen2023} have been favored by researchers. The idea of such approaches is to divide the states of agents into two sub-states with one containing insignificant information for communication with other agents and the other containing sensitive information only for internal information exchange. Another extension of privacy-preserving consensus is dynamics-based approaches \cite{Cheng2024a,Gao2018,Gao2022,Liu2022}, which is also the focus of this work. An important benefit of such approaches is that no trade-off exists between privacy and consensus performances, and they are easy to implement in conjunction with techniques like homomorphic encryption, differential privacy, etc. In contrast to state-decomposition strategy, dynamics-based approaches have a simpler structure and seem easier to much understand and implement. Note that some of the above privacy-preserving strategies have also been recently applied in decentralized learning literature \cite{Wei2024,Gade2018a,Cheng2024b,Han2022a,Gao2023,Lin2024a,Cheng2024c}.

\subsection{Main Contributions}
In this paper, our work contributes to enrich the dynamic-based privacy-preserving methods over unbalanced directed networks. Specifically, the contributions contain the points listed next.

\begin{enumerate}[i)]
\item{Based on the conventional push-sum structure, we design a novel push-sum average consensus method enabling privacy preservation. Specifically, during the initial several iterations, we ensure privacy preservation by carefully embedding randomness in mixing weights to confuse communications and introducing an extra auxiliary parameter to mask the state-updated dynamics. As well, to ensure consensus accuracy, exploiting the intrinsic robustness of consensus dynamics to cope with uncertain changes in information exchanges, we carefully redesign the push-sum protocol so that the ``total mass'' of the system is invariant in the presence of embedded randomness.}
\item{We provide a formal and rigorous analysis of convergence rate. Specifically, our analysis consists two parts. One is to analyze the consensus performance of the initial several iterations with randomness embedded, and the other is to analyze that of remaining randomness-free dynamics, which has the same structure as the conventional push-sum method \cite{Kempe2003,Benezit2010,Hadjicostis2018}. Our analysis exploits the properties of the mixing matrix product and norm relations to build consensus contractions of each dynamic. The result shows that the designed algorithm attains a linear convergence rate and explicitly captures the effect of mixing matrix and network connectivity structure on convergence rate.}
\item{Relaxing the privacy notion of considering only exact initial values in \cite{Manitara2013,Liu2005,Han2009,Cao2013}, we present two new privacy notions for honest-but-curious attacks and eavesdropping attacks (see Definition 3), respectively, where the basic idea is that the attacker has an infinite number of uncertainties in the estimation of the initial value through the available information. The privacy notions are more generalized in the context that the attacker is not only unable to determine the exact initial value but also the valid range of the initial value.}
\end{enumerate}

\emph{Notations:} $\mathbb{N}$ and $\mathbb{R}$ are the natural and real number sets, respectively. $\mathbf{0}$, $\mathbf{1}$, and $\mathbf{I}$ represent all-zero vector, all-one vector, and identity matrix, respectively, whose dimensions are clear from context. $\mathbf{A}=[ A_{ij} ] _{N\times N}$ represents an $N\times N$-dimensional matrix whose $ij$-th element is $A_{ij}$. $[ \cdot ] ^{\top}$ denotes the transpose of $[ \cdot ]$. The symbol ``$\setminus$'' stands for set subtraction. The symbol $| \cdot |$ is applied to the set to represent the cardinality and to the scalar value to represent absolute value. The $\ell _2$-norm (resp. $\ell _1$-norm) is signified by $\lVert \cdot \rVert$ (resp. $\lVert \cdot \rVert_1$).

\section{Preliminaries}
We recall several important properties and concepts associated with the graph theory, conventional push-sum protocol, and privacy.

\subsection{Graph Theory}
Consider a network consisting of $N$ agents and it is modeled as a digraph $\mathcal{G}=( \mathcal{V}, \mathcal{E} )$, where $\mathcal{V}=\{ 1,\cdots ,N \}$ is the agent set, and $\mathcal{E}$ is the edge set which comprises of pairs of agents and characterizes the interactions between agents, i.e., agent $i$ affects the dynamics of agent $j$ if a directed line from $i$ to $j$ exists, expressed as $( i,j ) \in \mathcal{E}$. Moreover, let $( i,i ) \notin \mathcal{E}$ for any $i\in \mathcal{V}$, i.e., no self-loop exists in digraph. Let $\mathcal{N}_{i}^{\text{in}}=\{ j| ( j, i ) \in \mathcal{E} \}$ and $\mathcal{N}_{i}^{\text{out}}=\{ j| ( i, j ) \in \mathcal{E} \}$ be the in-neighbor and out-neighbor sets of agent $i$, respectively. Accordingly, $D_{i}^{\text{in}}=| \mathcal{N}_{i}^{\text{in}} |$ and $D_{i}^{\text{out}}=| \mathcal{N}_{i}^{\text{out}} |$ denote the in-degree and out-degree, respectively. For any $i,j\in \mathcal{V}$, a trail from $i$ to $j$ is a chain of consecutively directed lines. The digraph $\mathcal{G}$ is \emph{strongly connected} if at least one trail lies between any pair of agents. The associated incidence matrix $\mathbf{R}=[ R_{i\varepsilon _j} ] _{N\times | \mathcal{E} |}$ of $\mathcal{G}$ is given by
\begin{flalign}
\nonumber R_{ie}=\begin{cases}
	1,&		\text{if the starting point of the}\,\,e\text{-th}\,\,\text{edge}\,\,(i,j)\,\,\text{is}\,\, i ;\\
	-1,&		\text{if the starting point of the}\,\,e\text{-th}\,\,\text{edge}\,\,(i,j)\,\,\text{is}\,\, j ;\\
	0,&		\text{otherwise}.\\
\end{cases}
\end{flalign}
One can know that the sum of each column of $\mathbf{R}$ is zero, i.e., $\sum\nolimits_{i=1}^N{R_{il}}=0$ for any $l \in [ 1,| \varepsilon | ]$. The mixing matrix $\mathbf{C}=[ C_{ij} ] _{N\times N}$ associated with $\mathcal{G}$ is defined as: $C_{ji}>0$ if $j\in \mathcal{N}_{i}^{\text{out}}\cup \{ i \}$ and $C_{ji}=0$ otherwise.

\begin{definition}\label{D1}(Sum-one condition:)
For an arbitrary matrix $\mathbf{A}=[ A_{ij} ] _{N\times N}$, if $\sum_{i=1}^N{A_{ij}}=1$ for all $j \in \mathcal{V}$, then $\mathbf{A}$ is column-stochastic. We claim that each column of $\mathbf{A}$ satisfies the sum-one condition.
\end{definition}

\begin{assumption}\label{A1}
The directed network $\mathcal{G}=( \mathcal{V}, \mathcal{E} )$ is strongly connected, and it holds $| \mathcal{V} |=N>2$. Each column of the mixing matrix $\mathbf{C}$ satisfies the sum-one condition.
\end{assumption}

\subsection{Conventional Push-Sum Method}
\begin{algorithm}[htb]
    \renewcommand{\thealgorithm}{1}
	\caption{Push-sum method}
	\label{alg:1}
	\begin{algorithmic}[1]
		\STATE \textbf{Input:} Initial states $x_i( 0 ) =z_i( 0 ) =x_{i}^{0} \in \mathbb{R}$ and $y_i( 0 ) =1$ for any $i\in \mathcal{V}$. The mixing weight $\mathbf{C}=[ C_{ij} ] _{N\times N}$ associated with $\mathcal{G}$.
		\FOR{$k=0,1,\cdots$}
        \FOR{$i=1,\cdots, n$ in parallel}
        \STATE Agent $i$ sends the computed $C_{li}x_i( k )$ and $C_{li}y_i( k )$ to $l\in \mathcal{N}_{i}^{\text{out}}$.
        \STATE Agent $i$ uses $C_{ij}x_j( k )$ and $C_{ij}y_j( k )$ received from $j\in \mathcal{N}_{i}^{\text{in}}$ to update $x_i$ and $y_i$ as follows:
        \begin{flalign}
        \label{Eq:1} &x_i( k+1 ) =\sum_{j\in \mathcal{N}_{i}^{\text{in}}\cup \{ i \}}{C_{ij}x_j( k )},\tag{1}\\
        \label{Eq:2} &y_i( k+1 ) =\sum_{j\in \mathcal{N}_{i}^{\text{in}}\cup \{ i \}}{C_{ij}y_j( k )},\tag{2}
        \end{flalign}
        \STATE Agent $i$ computes $z_i( k+1 ) =x_i( k+1 ) /y_i( k+1 )$.
        \STATE Until a stopping criteria is satisfied, e.g., agent $i$ stops if $| z_i( k+1 ) -\bar{x}^0 |<\epsilon$ for some predefined $\epsilon >0$, where $\bar{x}^0 \triangleq \sum\nolimits_{j=1}^N{x_j( 0 )}/N$.
        \ENDFOR
        \ENDFOR
    \end{algorithmic}
\end{algorithm}

Regarding the investigation of average consensus, the push-sum algorithm \cite{Kempe2003,Benezit2010,Hadjicostis2018} is a well-established protocol, which is summarized in Algorithm 1. All agents simultaneously update two variable states: $x_i( k )$ and $y_i( k )$, and the \textit{sensitive information} of agent $i$ is the initial value $x_i( 0 )$. Define $\mathbf{x}( k ) =[ x_1( k ) ,\cdots ,x_N( k ) ] ^{\top}$, $\mathbf{y}( k ) =[ y_1( k ) ,\cdots ,y_N( k ) ] ^{\top}$, and $\mathbf{C}=[ C_{ij} ] _{N\times N}$. We can rewrite \eqref{Eq:1} and \eqref{Eq:2} in a compact form as follows:
\begin{flalign}
\label{Eq:3} &\mathbf{x}( k+1 ) =\mathbf{Cx}( k ), \tag{3}
\\
\label{Eq:4} &\mathbf{y}( k+1 ) =\mathbf{Cy}( k ), \tag{4}
\end{flalign}
initialized with $\mathbf{x}( 0 ) =[ x_{1}^{0},\cdots ,x_{N}^{0} ] ^{\top}$ and $\mathbf{y}( 0 ) =\mathbf{1}$.

Under Assumption \ref{A1}, $\mathbf{C}^k$ converges to rank-one matrix at an exponential rate \cite{Seneta1973,Fill1991}. Let $\mathbf{C}^{\infty}$ be the infinite power of matrix $\mathbf{C}$, i.e., $\mathbf{C}^{\infty}=\lim _{k\rightarrow \infty}\,\,\mathbf{C}^k$. Applying the Perron-Frobenius theorem \cite{Horn2012} gives $\mathbf{C}^{\infty}=\boldsymbol{\pi }\mathbf{1}^{\top}$, where $\boldsymbol{\pi }=[ \pi _1,\cdots ,\pi _N ] ^{\top}$. Recursively calculating \eqref{Eq:3} and \eqref{Eq:4} yields:
\begin{flalign}
\nonumber \mathbf{x}( k ) =\mathbf{C}^k\mathbf{x}( 0 ),\,\,\mathbf{y}( k ) =\mathbf{C}^k\mathbf{y}( 0 ).
\end{flalign}
Then, it follows that
\begin{flalign}
\label{Eq:5} \underset{k\rightarrow \infty}{\lim}\,\,z_i( k ) =\underset{k\rightarrow \infty}{\lim}\,\,\frac{x_i( k )}{y_i( k )}=\frac{[ \mathbf{C}^{\infty}\mathbf{x}( 0 ) ] _i}{[ \mathbf{C}^{\infty}\mathbf{y}( 0 ) ] _i}=\frac{\pi _i\sum\nolimits_{j=1}^N{x_j( 0 )}}{\pi _i\sum\nolimits_{j=1}^N{y_j( 0 )}}=\frac{\sum\nolimits_{j=1}^N{x_j( 0 )}}{N}, \tag{5}
\end{flalign}
where $[ \cdot ] _i$ means the $i$-th element of $[ \cdot ]$. Thus, the ratio $z_i( k )$ gradually reaches to $\bar{x}^0$. More details of the analysis can be found in \cite{Kempe2003}.

\subsection{Privacy Concern}
We introduce two prevalent attack types, namely, honest-but-curious attacks and eavesdropping attacks. Then, we explain that Algorithm 1 fails to preserve privacy due to the explicit sharing of state variables.

\begin{definition}
An honest-but-curious attack is an attack in which some agents, who follow the state-update protocols properly, try to infer the initial values of other agents by using the received information.
\end{definition}

\begin{definition}
An eavesdropping attack is an attack in which an external eavesdropper is able to capture all sharing information by wiretapping communication channels so as to infer the private information about sending agents.
\end{definition}

In general, in terms of information leakage, an eavesdropping attack is more devastating than an honest-but-curious attack as it can capture all transmitted information, while the latter can only access the received information. Yet, the latter has the advantage that the initial values $\{ x_{j}^{0} \}$ of all honest-but-curious agents $j$ are accessible, which are unavailable to the external eavesdroppers.

For the average consensus, the sensitive information to be protected is the initial value $x_i( 0 )$, $i \in \mathcal{V}$. At the first iteration, agent $i$ will send the computed values $C_{ji}x_i( 0 )$ and $C_{ji}y_i( 0 )$ to all of its out-neighbors $j\in \mathcal{N}_{i}^{\text{out}}$. Then, the initial value $x_i( 0 )$ is uniquely inferable by the honest-but-curious agent $j$ using $x_i( 0 ) =\frac{C_{ij}x_i( 0 )}{C_{ij}y_i( 0 )}$ and $y_i( 0 ) =1$. Therefore, the honest-but-curious agents are always able to infer the sensitive information of its in-neighbors. Likewise, one can readily check that external eavesdroppers are also able to easily infer sensitive information about all agents. Therefore, the privacy concern is not addressed in the conventional push-sum method. In this work, we try to study the privacy concern and develop a privacy-preserving version of Algorithm 1 to achieve exact average consensus.

\subsection{Performance Metric}
Our task is to propose an average consensus algorithm that can achieve exact convergence while guaranteeing privacy security. According to the above discussion, the following two requirements for privacy-preserving push-sum algorithms must be satisfied.
\begin{enumerate}[i)]
\item{Exact output: After the last iteration of the algorithm, each agent should converge to the average consensus point $\bar{x}^0$.}
\item{Privacy preservation: During the entire algorithm implementation, the private information, i.e., the initial value $x_{i}^{0}$, of each legitimate agent $i$ should be preserved against both honest-but-curious and eavesdropping attacks. }
\end{enumerate}

In order to respond to the above two requirements, two metrics are required to quantify them.

\textbf{Output metric:} To measure the accuracy of the output, we adopt the consensus error $\lVert \mathbf{z}( k ) -\bar{x}^0\mathbf{1} \rVert$. The algorithm achieves exact consensus if $\lim _{k\rightarrow \infty}\lVert \mathbf{z}( k ) -\bar{x}^0\mathbf{1} \rVert =0$. Furthermore, the algorithm is said to be \emph{elegant} if $\lVert \mathbf{z}( k ) -\bar{x}^0\mathbf{1} \rVert =\mathcal{O}( \rho ^k )$, $\rho \in ( 0,1 )$.

\textbf{Privacy metric:} For the honest-but-curious attacks, we consider the presence of some honest-but-curious agents $\mathcal{H}$. The accessible information set of $\mathcal{H}$ is represented as $\mathcal{I}_h( k ) =\{ \mathcal{I}_j( k ) | j\in \mathcal{H} \}$, where $\mathcal{I}_j( k )$ represents the information available to agent $j\in \mathcal{H}$ at iteration $k$. Given a moment $k'\in \mathbb{N}$, the access information of agents $\mathcal{H}$ in time period $0-k$ is $\mathcal{I}_h( 0:k' ) =\cup _{0\le k\le k'}\mathcal{I}_h( k )$. For any information sequence $\mathcal{I}_h( 0:k' )$, define $\mathcal{S}_{0}^{i}$ as the set of all possible initial values at the legitimate agent $i$, where all initial values leave the information accessed by agents $\mathcal{H}$ unchanged. That is to say, there exist any two initial values $x_{i}^{0},\tilde{x}_{i}^{0}\in \mathcal{S}_{0}^{i}$ with $x_{i}^{0}\ne \tilde{x}_{i}^{0}$ such that $\tilde{\mathcal{I}}_h( 0:k' ) =\mathcal{I}_h( 0:k' )$. The diameter of $\mathcal{S}_{0}^{i}$ is defined as
\begin{flalign}
\nonumber \mathbf{D}( \mathcal{S}_{0}^{i} ) =\underset{x_i( 0 ) ,\tilde{x}_i( 0 ) \in \mathcal{S}_{0}^{i}}{\text{sup}}| x_i( 0 ) -\tilde{x}_i( 0 ) |.
\end{flalign}

For the eavesdropping attacks, we consider the presence of an external eavesdropper whose available information is denoted as $\mathcal{I}_e( k )$, $k\in \mathbb{N}$. Let $\mathcal{I}_e( 0:k' ) =\cup _{0\le k\le k'}\mathcal{I}_e( k )$. Similar to the honest-but-curious attacks, we define $\mathcal{S}_0$ as the set of all possible initial values for all agents, where all initial values leave the information accessed by an external eavesdropper unchanged. That is, there exist $\mathbf{x}( 0 ), \mathbf{\tilde{x}}( 0 ) \in \mathcal{S}_0$ with $\mathbf{x}( 0 ) \ne \mathbf{\tilde{x}}( 0 )$ such that $\mathcal{I}_e( k ) =\tilde{\mathcal{I}}_e( k )$. In addition, the diameter of $\mathcal{S}_0$ is given as
\begin{flalign}
\nonumber \mathbf{D}( \mathcal{S}_0 ) =\underset{\mathbf{x}( 0 ) ,\mathbf{\tilde{x}}( 0 ) \in \mathcal{S}_0}{\text{sup}}\lVert \mathbf{x}( 0 ) -\mathbf{\tilde{x}}( 0 ) \rVert.
\end{flalign}

For the honest-but-curious and eavesdropping attacks, we use $\mathbf{D}( \mathcal{S}_{0}^{i} )$ for all legitimate agents $i\in \mathcal{V}\setminus \mathcal{H}$ and $\mathbf{D}( \mathcal{S}_0 )$ for all agents to measure the individual privacy and algorithm-level confidentiality, respectively. For more details, see the definition below.

\begin{definition}
The algorithm is said to be elegant in terms of privacy preservation, if $\mathbf{D}( \mathcal{S}_{0}^{i} ) =\infty$ or $\mathbf{D}( \mathcal{S}_0 ) = \infty$ for any information sequence $\mathcal{I}_h( 0:k' )$ or $\mathcal{I}_e( 0:k' )$, $k'\in \mathbb{N}$, respectively.
\end{definition}

The privacy concept outlined in Definition 4 shares similarities with the uncertainty-based privacy concept presented in \cite{Lu2020}, which derives inspiration from the $l$-diversity principle \cite{Machanavajjhala2007}. Within the $l$-diversity framework, the variety of any private information is gauged by the number of disparate estimates produced for the information. The higher this diversity, the more ambiguous the associated private information becomes. In our setting, the privacy information is the initial value $x_{i}^{0}$ (resp. $\mathbf{x}( 0 )$), whose diversity is measured by the diameter $\mathbf{D}( \mathcal{S}_{0}^{i} )$ (resp. $\mathbf{D}( \mathcal{S}_0 )$). Larger diameters imply greater uncertainty in the estimation of the initial values.

\begin{remark}
Note that Definition 4 indicates that attackers cannot uniquely determine an exact value or even a valuable range of $x_{i}^{0}$, and hence is more stringent than the notion defined in \cite{Manitara2013,Liu2005,Han2009,Cao2013}, which only considers the privacy information not to be exactly inferred.
\end{remark}

\section{Privacy-Preserving Push-Sum Algorithm}
Based on the discussion of Algorithm 1, one can know that adopting the same weight $C_{ij}$ for both $C_{ij}x_i( 0 )$ and $C_{ij}y_i( 0 )$ cause privacy (i.e., initial values) leakage. To solve the issue, a dynamics-based weight generation mechanism is developed in \cite{Gao2018}, whose details are outlined in Protocol 1.

\floatname{algorithm}{Protocol}
\begin{algorithm}[htb]
    \renewcommand{\thealgorithm}{1}
	\caption{Weight generation mechanism}
	\label{Pro:1}
	\begin{algorithmic}[1]
    \STATE \textbf{Required parameters:} Parameters $K\in \mathbb{N}$ and $\eta \in ( 0,1 )$ are known to each agent.
    \STATE Two sets of tailored mixing weights associated with any edge $( j,i ) \!\in\! \mathcal{E}$ are generated. Specifically, when $k\!\le \!K$, two groups of mixing weights $\{ C_{ji}^{1}( k ) \! \in \! \mathbb{R}|\! \,\,j\!\in \!\mathcal{N}_{i}^{\text{out}}\!\cup\! \{ i \}  \}$ and $\{ C_{ji}^{2}( k ) \!\in \!\mathbb{R}| \!\,\,j\!\in\! \mathcal{N}_{i}^{\text{out}}\!\cup\! \{ i \} \}$ associated with agent $i$ are generated, which satisfy $\sum\nolimits_{j=1}^N{C_{ji}^{1}( k )}=1$ and $\sum\nolimits_{j=1}^N{C_{ji}^{2}( k )}=1$; when $k\!> \!K$, only one group of mixing weights $\{ C_{ji}( k )=C_{ji}^{1}( k ) =C_{ji}^{2}( k ) \in ( \eta ,1 ) | \,\,j\in \mathcal{N}_{i}^{\text{out}}\cup \{ i \}  \}$, satisfying $\sum\nolimits_{j=1}^N{C_{ji}( k )}=1$, is generated. Note that $\{ C_{ji}^{1}( k )\}$ and $\{ C_{ji}^{2}( k )\}$ are mixed in $x_i$ and $y_i$, respectively. Moreover, agent $i$ always sets $C_{ji}^{1}( k ) =0$ and $C_{ji}^{2}( k ) =0$ for $j\notin \mathcal{N}_{i}^{\text{out}}\cup \{ i \}$.
    \end{algorithmic}
\end{algorithm}

The main idea of the dynamics-based mechanism is to confuse the state variables of the agents by injecting randomness into the mixing matrix in the initial few iterations. Fig. 1 briefly depicts the basic process. Obviously, the dynamics-based protocol has two stages. The first stage is from iteration $k=0$ to $k=K$, which can be regarded as a re-initialization operation on the initial value. This stage is key to privacy protection. The second stage is from $k=K+1$ to $k=\infty$, which can be viewed as the normal executions of the conventional push-sum method. This stage is key to ensuring convergence.

\begin{figure}[htb]
\centering
\includegraphics[width=14cm]{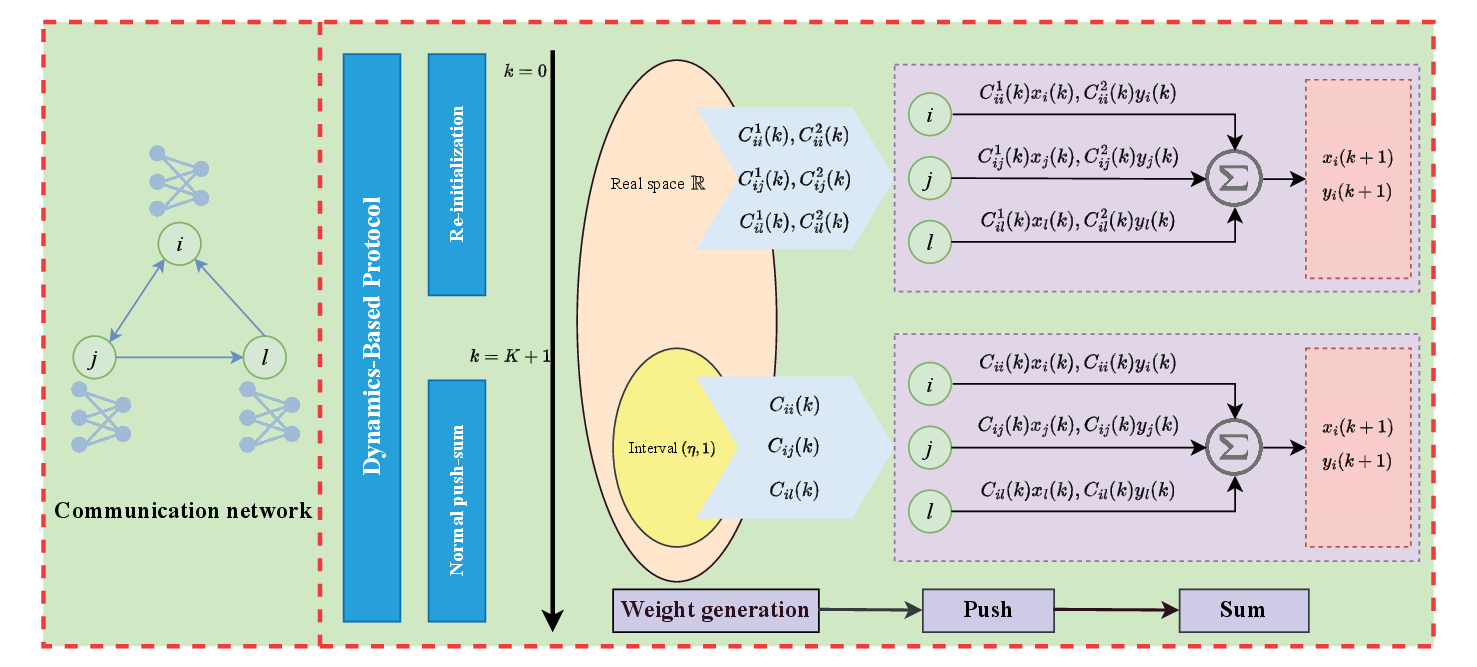}
\caption{Dynamics-based protocol: A brief computation process from the view of agent $i$ over a simple $3$-agent digraph.}
\label{fig:1}
\centering
\end{figure}

The dynamics-based method has been proved to reach an exact consensus point, and the sensitive information of legitimate agents is not inferred by honest-but-curious attackers in \cite{Gao2018}. However, there are three significant challenges that have not been addressed: I) In the initial $K$ iterations, although each weight is arbitrary, the sum-one condition still imposes a constraint on the weight setting; II) The method cannot protect sensitive information from external eavesdropping attackers; II) Only asymptotic convergence of algorithms is discussed in most of the average consensus literature \cite{Wang2019,Chen2023,Cheng2024a,Gao2018,Gao2022,Liu2022}, and analysis of the speed of convergence is rare.

To solve the above issues, we carefully redesign the push-sum rule to address I) and II), and III) is tackled in Section IV. From Protocol 1, one knows that the dynamics-based method mainly operates on the first $K$ iterations to preserve the privacy information. Specifically, the update rule of the $x$-variable is given as
\begin{flalign}
\nonumber x_i( k+1 ) =\sum_{j\in \mathcal{N}_{i}^{\text{in}}\cup \{ i \}}{C_{ij}^{1}( k )x_j( k )}, \,\,k \le K,
\end{flalign}
where $C_{ij}^{1}( k )$ is generated from Protocol 1. Note that the sum-one condition is used to ensure that the sum of all variables at each iteration $k \le K$ is invariant, that is,
\begin{flalign}
\label{Eq:6} \sum_{i=1}^N{x_i( k+1 )}=\sum_{i=1}^N{x_i( k )}. \tag{6}
\end{flalign}
Thus, if we wish to circumvent the sum-one constraint, the new update rule must make \eqref{Eq:6} hold. Specifically, we take advantage of the fact that the amount of messages sent and received is equal for the entire system (i.e., the total mass of the system is fixed) and modify the update of the $x$-variable as
\begin{flalign}
\label{Eq:7} x_i( k+1 ) =x_i( k ) +\varXi_i ( k )  \tag{7}
\end{flalign}
with \[\varXi_i ( k )\! \triangleq \!\sum_{j\in \mathcal{N}_{i}^{\text{in}}}{C_{ij}^{1}( k ) x_j( k )}\!-\!\sum_{j\in \mathcal{N}_{i}^{\text{out}}}{C_{ji}^{1}( k ) x_i( k )},\] where $C_{ij}^{1}( k )$ can take any value in $\mathbb{R}$ (the sum-one condition is not required). One verifies that $\sum\nolimits_{i=1}^N{\varXi_i ( k )}=0$. Obviously, summing $x_i( k+1 )$ in \eqref{Eq:7} over $i=1,\cdots ,N$ yields \eqref{Eq:6}. However, the update rule \eqref{Eq:7} is valid for honest-but-curious attacks and still ineffective for eavesdropping attacks, see Corollary 2. Thus, we further introduce an auxiliary parameter $\sigma ( k ) \in \mathbb{R}$ for $k\le K$, which is public information known for all agents, but not to the external eavesdropper. Details of our method are summarized in Algorithm \ref{alg:2}.

\floatname{algorithm}{Algorithm}
\begin{algorithm}[!htb]
    \renewcommand{\thealgorithm}{2}
	\caption{Secure average consensus algorithm}
	\label{alg:2}
	\begin{algorithmic}[1]
		\STATE \textbf{Input:} Initial states $x_i( 0 ) =z_i( 0 ) =x_{i}^{0}$ and $y_i( 0 ) =1$ for $i\in \mathcal{V}$; Parameters $K\in \mathbb{N}$, $\sigma ( k ) \in \mathbb{R}$ for $k\in \mathbb{N}$, and $\eta \in ( 0,1 )$; Network $\mathcal{G}$.
        \STATE \textbf{Weight generation:} Two sets of random mixing weights associated with any edge $( j,i ) \in \mathcal{E}$ are generated. For the variable $y_i( k )$ at any $k\in \mathbb{N}$, a group of mixing weights $\{ C_{ji}^{2}( k ) \in ( \eta ,1 ) | j\in \mathcal{N}_{i}^{\text{out}}\cup \{ i \} \}$ are generated, which satisfy $\sum\nolimits_{j\in \mathcal{N}_{i}^{\text{out}}\cup \{ i \}}^{\,\,}{C_{ji}^{2}( k )}=1$. For the variable $x_i( k )$, if $k\le K$, a group of mixing weights $\{ C_{ji}^{1}( k ) \in \mathbb{R}| \,\,j\in \mathcal{N}_{i}^{\text{out}}\cup \{ i \} \}$ are generated; Otherwise, a group of mixing weights $\{ C_{ji}^{1}( k ) =C_{ji}^{2}( k ) \in ( \eta ,1 ) | j\in \mathcal{N}_{i}^{\text{out}}\cup \{ i \} \}$ are generated. Moreover, agent $i$ always sets $C_{ji}^{1}( k ) =0$ and $C_{ji}^{2}( k ) =0$ for $j\notin \mathcal{N}_{i}^{\text{out}}\cup \{ i \}$.
		\FOR{$k=0,1,\cdots$}
        \FOR{$i=1,\cdots,n$ in parallel}
        \STATE Agent $i$ sends the computed $C_{li}^{1}( k ) x_i( k )$ and $C_{li}^{2}( k ) y_i( k )$ to $ l \in \mathcal{N}_{i}^{\text{out}}$.
        \STATE Agent $i$ uses $C_{ij}^{1}( k )x_j( k )$ and $C_{ij}^{2}( k )y_j( k )$ received from $j\in \mathcal{N}_{i}^{\text{in}}$ to update $x_i$ and $y_i$ as follows:
        \begin{flalign}
        \label{Eq:8}&x_i( k+1 ) \!=\!\begin{cases}
	x_i( k ) \!+\!\sigma ( k ) \varXi_i ( k ) ,&		\text{if}\,\,k\le K;\\
	\underset{j\in \mathcal{N}_{i}^{\text{in}}\cup \{ i \}}{\sum}C_{ij}^{1}( k ) x_j( k ),&		\text{if}\,\,k\ge K+1.\\
\end{cases} \tag{8}
\\
\label{Eq:9} &y_i( k+1 )=\underset{j\in \mathcal{N}_{i}^{\text{in}}\cup \{ i \}}{\sum}C_{ij}^{2}( k ) y_j( k ), k\ge 0. \tag{9}
        \end{flalign}
        \STATE Agent $i$ computes $z_i( k+1 ) =x_i( k+1 ) /y_i( k+1 )$.
        \STATE Until a stopping criteria is satisfied, e.g., agent $i$ stops if $| z_i( k+1 ) -\bar{x}^0 |<\epsilon$ for some predefined $\epsilon >0$.
        \ENDFOR
        \ENDFOR
    \end{algorithmic}
\end{algorithm}

\begin{remark}
Note that we mainly embed randomness for $\mathbf{C}_1( k )$ in the first $K$ iterations and do not consider $\mathbf{C}_2( k )$. Since embedding randomness for $\mathbf{C}_1( k )$ alone can guarantee that $\mathbf{C}_1( k ) \ne \mathbf{C}_2( k )$ for $k \le K$, and the auxiliary variable $y$ does not contain privacy information, so there is no need to embed randomness for $\mathbf{C}_2( k )$ either. Of course, if embedding randomness for $\mathbf{C}_2( k )$ is necessary, the update of the $y$-variable in \eqref{Eq:9} is formulated as:
\begin{flalign}
\nonumber y_i( k+1 )=y_i( k ) \!+\!\sigma^{'}( k ) \Big( \sum_{j\in \mathcal{N}_{i}^{\text{in}}}{C_{ij}^{2}( k )y_j( k )}\!-\!\sum_{j\in \mathcal{N}_{i}^{\text{in}}}{C_{ij}^{2}( k )y_j( k )} \Big),
\end{flalign}
where $\sigma^{'}( k )$ and $C_{ij}^{2}( k )$ are generated in a similar way as $\sigma( k )$ and $C_{ij}^{1}( k )$ of Algorithm 2.
\end{remark}

\section{Convergence Analysis}
Following Algorithm \ref{alg:2}, it holds from the dynamics \eqref{Eq:8}-\eqref{Eq:9} that
\begin{flalign}
\label{Eq:10} &\mathbf{x}( k+1 ) =\mathbf{C}_1( k ) \mathbf{x}( k ), k\ge K, \tag{10}
\\
\label{Eq:11} &\mathbf{y}( k+1 ) =\mathbf{C}_2( k ) \mathbf{y}( k ), k\ge 0, \tag{11}
\end{flalign}
where $\mathbf{C}_1( k ) =[ C_{ij}^{1}( k ) ] _{N\times N}$ and $\mathbf{C}_2( k ) =[ C_{ij}^{2}( k ) ] _{N\times N}$. It is clear from the settings of Algorithm \ref{alg:2} that: i) $\mathbf{C}_1( k )$ and $\mathbf{C}_2( k )$ are time-varying and column-stochastic; and ii) $\mathbf{C}_1( k ) =\mathbf{C}_2( k )$ for $k\ge K$.

Define $\mathbf{\Phi }_1( k:s ) =\mathbf{C}_1( k ) \cdots \mathbf{C}_1( s )$ and $\mathbf{\Phi }_2( k:s ) =\mathbf{C}_2( k ) \cdots \mathbf{C}_2( s )$ for $k\ge s\ge 0$. Particularly, $\mathbf{\Phi }_1( k:k ) =\mathbf{C}_1( k )$ and $\mathbf{\Phi }_2( k:k ) =\mathbf{C}_2( k )$. Recursively computing \eqref{Eq:10} and \eqref{Eq:11}, we can obtain
\begin{flalign}
\label{Eq:12} &\mathbf{x}( k+1 ) =\mathbf{\Phi }_1( k:K+1 ) \mathbf{x}( K+1 ), k\ge K+1, \tag{12}
\\
\label{Eq:13} &\mathbf{y}( k+1 ) =\mathbf{\Phi }_2( k:0 ) \mathbf{y}( 0 ), k\ge 0, \tag{13}
\end{flalign}
where it holds $\mathbf{\Phi }_1( k:K+1 ) =\mathbf{\Phi }_2( k:K+1 )$ for $k\ge K+1$.
Then, it follows that
\begin{flalign}
\label{Eq:14} &\mathbf{1}^{\top}\mathbf{x}( k+1 ) =\mathbf{1}^{\top}\mathbf{x}( K+1 ), k\ge K+1, \tag{14}
\\
\label{Eq:15} &\mathbf{1}^{\top}\mathbf{y}( k+1 ) =\mathbf{1}^{\top}\mathbf{y}( 0 ) =N, k\ge 0, \tag{15}
\end{flalign}
where we use the column stochasticities of $\mathbf{\Phi }_1( k:K+1 )$ and $\mathbf{\Phi }_2( k:0 )$. For the first $K$ dynamics of $x_i$ in \eqref{Eq:8}, using the fact that $\sum\nolimits_{i=1}^N{\varXi_i ( k )}=0$ gives
\begin{flalign}
\nonumber \mathbf{1}^{\top}\mathbf{x}( k+1 ) =&\sum_{i=1}^N{x_i( k+1 )}=\sum_{i=1}^N{( x_i( k ) +\sigma ( k ) \varXi_i ( k ) )}
\\
\label{Eq:16} \,\,             =&\sum_{i=1}^N{x_i( k )}=\mathbf{1}^{\top}\mathbf{x}( k ) =\mathbf{1}^{\top}\mathbf{x}( 0 ), \tag{16}
\end{flalign}
which matches the relation \eqref{Eq:6}. Combining \eqref{Eq:14} and \eqref{Eq:16} gives
\begin{flalign}
\label{Eq:17} \mathbf{1}^{\top}\mathbf{x}( k+1 ) =\mathbf{1}^{\top}\mathbf{x}( 0 ), k\ge 0. \tag{17}
\end{flalign}

Note that the dynamics of Algorithm \ref{alg:2} for iterations $k\ge K$ are analogous to the conventional push-sum method. Considering \eqref{Eq:17} in depth, it can be seen that the injected randomness of the first $K$ dynamics has no impact on the consensus performance. Next we show that Algorithm \ref{alg:2} can guarantee a linear convergence rate. Let $\mathbf{z}( k ) =[ z_1( k ) ,\cdots ,z_N( k ) ] ^{\top}$.

\begin{theorem}
Let $\{ ( z_i( k ) ) _{i=1}^{N} \} _{k\in \mathbb{N}}$ be the sequence generated by Algorithm \ref{alg:2}, and the network $\mathcal{G}$ satisfies Assumption \ref{A1}. Then, it holds, for all $k\in \mathbb{N}$, \[ \lVert \mathbf{z}( k ) -\bar{x}^0\mathbf{1} \rVert \le c\rho ^k, \] where $\rho =( 1-\eta ^{N-1} ) ^{\frac{1}{N-1}}$, and $c$ is a constant given as
\[ c= \max \bigg\{\!\!\!\! \begin{array}{c}
	c_1,( c_2+c_3 ) \lVert \mathbf{x}( 0 ) \rVert _1,( c_2\rho ^{-1}+c_3 ) \lVert \mathbf{x}( 1 ) \rVert _1,\\
	\cdots ,( c_2\rho ^{-K-1}+c_3 ) \lVert \mathbf{x}( K+1 ) \rVert _1\\
\end{array}\!\!\!\! \bigg\}, \]
where $c_1=2\sqrt{N}c_0\lVert \mathbf{x}( K+1 ) \rVert _1\eta ^{-N}\rho ^{-K-2}$, $c_2=2\sqrt{N}\eta ^{-N}-( N-1 ) /\sqrt{N}$ and $c_3=N^{-1/2}\eta ^{-N}c_0\rho ^{-1}$.
\begin{proof}
Details of the analysis are outlined in Appendix A.
\end{proof}
\end{theorem}

\begin{remark}
Theorem 1 indicates that Algorithm 1 can achieve an $\mathcal{O}( \rho ^k )$ convergence rate with $\rho =( 1-\eta ^{N-1} ) ^{\frac{1}{N-1}}$. Evidently, a smaller $\rho$ yields a better convergence rate. A straightforward way to obtain a smaller $\rho$ is to increase $\eta$. However, it is essential to be aware that $\eta$ cannot be close to $1$ arbitrarily due to the nonnegativity and column stochasticity of the mixing matrix for $k \ge K+1$. To satisfy the weight generation mechanism in Algorithm 2, it holds $0\le \eta \le 1/( \max _i {D_{i}^{\text{out}}} + 1 )$.
\end{remark}

\section{Privacy Analysis}
We analyze that Algorithm 2 is resistant to both honest-but-curious and eavesdropping attacks.

\subsection{Performance Against Honest-but-curious Attacks}
For the honest-but-curious attacks, we make the following standard assumption.

\begin{assumption}\label{A2}
Consider a strongly connected network $\mathcal{G}$, where some colluding honest-but-curious nodes exist. We assume that each agent $i\in \mathcal{V}$ has at least one legitimate neighbor, i.e., $\mathcal{N}_{i}^{\text{out}}\cup \mathcal{N}_{i}^{\text{in}}\nsubseteq \mathcal{H}$.
\end{assumption}

\begin{remark}
Assumption 2 is common in security consensus literature \cite{Wang2019,Chen2023,Cheng2024a,Gao2018,Gao2022,Liu2022}. For example, in a real network, we hold an agent $i$. If all other agents in the network are untrustworthy, we can simply add another agent $l$ that we hold to the network so that it is externalized to agent $i$.
\end{remark}

\begin{theorem}
Under Assumptions \ref{A1}-\ref{A2}, the initial value $x_{i}^{0}$ of legitimate agent $i\in \mathcal{V}$ can be safely protected if $\mathcal{N}_{i}^{\text{out}}\cup \mathcal{N}_{i}^{\text{in}}\nsubseteq \mathcal{H}$ during the running of Algorithm 2.
\begin{proof}
Recalling the definition of privacy metric in Section II-D, it can be shown that the privacy of agent $i$ can be safely protected insofar as $\mathbf{D}( \mathcal{S}_{0}^{i} ) =\infty$. The available information to $\mathcal{H}$ is $\mathcal{I}_h=\{ \mathcal{I}_j| j\in \mathcal{H} \}$, where $\mathcal{I}_j$ denotes the information available to each individual $j\in \mathcal{H}$ given as
\begin{flalign}
\nonumber \mathcal{I}_j=&\{ \mathcal{I}_{j}^{\text{state}}( k ) \cup \mathcal{I}_{j}^{\text{send}}( k ) \cup \mathcal{I}_{j}^{\text{receive}}( k ) | k\ge 0 \}
\\
\nonumber \,\,\,\,    &\cup \{ \sigma ( k ) | 0\le k\le K \} \cup \{ y_m( 0 ) =1| m\in \mathcal{V} \}
\\
\nonumber \,\,     &\cup \{ C_{nj}^{1}( k ) ,C_{nj}^{2}( k ) | m\in \mathcal{V},k\ge 0 \}
\end{flalign}
with
\begin{flalign}
\nonumber &\mathcal{I}_{j}^{\text{state}}( k ) =\{ x_j( k ) ,y_j( k ) \}
\\
\nonumber &\mathcal{I}_{j}^{\text{send}}( k ) \!=\!\{ C_{nj}^{1}( k ) x_j( k ), C_{nj}^{2}( k ) y_j( k ) | n\in \mathcal{N}_{j}^{\text{out}}\cup \{ j \} \}
\\
\nonumber &\mathcal{I}_{j}^{\text{receive}}( k ) =\{ C_{jm}^{1}( k ) x_m( k ) , C_{jm}^{2}( k ) y_m( k ) | m\in \mathcal{N}_{j}^{\text{in}} \}.
\end{flalign}
To prove $\mathbf{D}( \mathcal{S}_{0}^{i} ) =\infty$, it suffices to show that agents in $\mathcal{H}$ fail to judge whether the initial value of agent $i$ is $x_{i}^{0}$ or $\tilde{x}_{i}^{0}=x_{i}^{0}+\delta$ where $\delta$ is an arbitrary value in $\mathbb{R}$ and $x_{i}^{0}, \tilde{x}_{i}^{0}\in \mathcal{S}_{0}^{i}$. Note that agents in $\mathcal{H}$ are only able to infer $x_{i}^{0}$ using $\mathcal{I}_h$. In other words, if the initial value $\tilde{x}_{i}^{0}=x_{i}^{0}+\delta$ makes the information $\tilde{\mathcal{I}}_{h}$ accessed by agents of $\mathcal{H}$ unchanged, i.e., $\tilde{\mathcal{I}}_{h}=\mathcal{I}_{h}$, then $\mathbf{D}( \mathcal{S}_{0}^{i} ) =\infty$. Hence, we only need to prove that there is $\tilde{\mathcal{I}}_{h}=\mathcal{I}_{h}$ under two different initial values $\tilde{x}_{i}^{0}$ and $x_{i}^{0}$.

Since $\mathcal{N}_{i}^{\text{out}}\cup \mathcal{N}_{i}^{\text{in}}\nsubseteq \mathcal{H}$, there exists at least one agent $l\in \mathcal{N}_{i}^{\text{out}}\cup \mathcal{N}_{i}^{\text{in}}\setminus \mathcal{H}$. Thus, some settings on initial values of agent $l$ and mixing weights associated with agent $l$ that satisfy the requirements in Algorithm \ref{alg:2} such that $\tilde{\mathcal{I}}_{h}=\mathcal{I}_{h}$ holds for any variant $\tilde{x}_{i}^{0}$. More specifically, the initial settings are given as
\begin{flalign}
\label{Eq:18} &\tilde{x}_{i}^{0}=x_{i}^{0}+\delta, \tilde{x}_{l}^{0}=x_{l}^{0}-\delta, \tilde{x}_{m}^{0}=x_{m}^{0}, m\in \mathcal{V}\setminus \{ i,l \}, \tag{18}
\end{flalign}
where $\delta$ is nonzero and does not equal either $-x_i( 0 )$ or $x_l( 0 )$. Apparently, such an initial value setting has no impact on the sum of the original initial values. Then, we properly choose the mixing weights such that $\tilde{\mathcal{I}}_{h}=\mathcal{I}_{h}$. Here, ``properly'' means the choosing mixing weights should obey the weight generation mechanism in Algorithm \ref{alg:2}. Our analysis will be continued in two cases, $l\in \mathcal{N}_{i}^{\text{out}}$ and $l\in \mathcal{N}_{i}^{\text{in}}$, respectively.

\noindent \textbf{Case I:} We consider $l\in \mathcal{N}_{i}^{\text{out}}$. One derives $\tilde{\mathcal{I}}_{h}=\mathcal{I}_{h}$ if the weights are set as
\begin{flalign}
\label{Eq:19a} &\tilde{C}_{mn}^{1}( 0 ) =C_{mn}^{1}( 0 ), m\in \mathcal{V}, n\in \mathcal{V}\setminus \{ i,l \}, \tag{19a}
\\
\label{Eq:19b} &\tilde{C}_{mi}^{1}( 0 ) =C_{mi}^{1}( 0 ) x_{i}^{0} /\tilde{x}_{i}^{0}, m\in \mathcal{V}\setminus \{ i,l \}, \tag{19b}
\\
\label{Eq:19c} &\tilde{C}_{li}^{1}( 0 ) =( \sigma ( 0 ) C_{li}^{1}( 0 ) x_{i}^{0} +\delta ) /\sigma ( 0 ) \tilde{x}_{i}^{0}, \tag{19c}
\\
\label{Eq:19d} &\tilde{C}_{ml}^{1}( 0 ) =C_{ml}^{1}( 0 ) x_{l}^{0} /\tilde{x}_{l}^{0}, m\in \mathcal{V}\setminus \{ l \}, \tag{19d}
\\
\label{Eq:19e} &\tilde{C}_{ii}^{1}( 0 ) ,\tilde{C}_{ll}^{1}( 0 ) \in \mathbb{R}, \tag{19e}
\\
\label{Eq:19f} &\tilde{C}_{mn}^{1}( k ) =C_{mn}^{1}( k ), m,n\in \mathcal{V}, k\ge 1, \tag{19f}
\\
\label{Eq:19g} &\tilde{C}_{mn}^{2}( k ) =C_{mn}^{2}( k ), m,n\in \mathcal{V}, k\ge 0. \tag{19g}
\end{flalign}

\noindent \textbf{Case II:} We consider $l\in \mathcal{N}_{i}^{\text{in}}$. One derives $\tilde{\mathcal{I}}_{h}=\mathcal{I}_{h}$ if the weights are set as
\begin{flalign}
\label{Eq:20a} &\tilde{C}_{mn}^{1}( 0 ) =C_{mn}^{1}( 0 ), m\in \mathcal{V}, n\in \mathcal{V}\setminus \{ i,l \}, \tag{20a}
\\
\label{Eq:20b} &\tilde{C}_{mi}^{1}( 0 ) =C_{mi}^{1}( 0 ) x_{i}^{0} /\tilde{x}_{i}^{0}, m\in \mathcal{V}\setminus \{ i \}, \tag{20b}
\\
\label{Eq:20c} &\tilde{C}_{ml}^{1}( 0 ) =C_{ml}^{1}( 0 ) x_{l}^{0} /\tilde{x}_{l}^{0}, m\in \mathcal{V}\setminus \{ i,l \}, \tag{20c}
\\
\label{Eq:20d} &\tilde{C}_{il}^{1}( 0 ) =( \sigma ( 0 ) C_{il}^{1}( 0 ) x_{l}^{0} -\delta ) /\sigma ( 0 ) \tilde{x}_{l}^{0}, \tag{20d}
\\
\label{Eq:20e} &\tilde{C}_{ii}^{1}( 0 ) ,\tilde{C}_{ll}^{1}( 0 ) \in \mathbb{R}, \tag{20e}
\\
\label{Eq:20f} &\tilde{C}_{mn}^{1}( k ) =C_{mn}^{1}( k ), m,n\in \mathcal{V}, k\ge 1, \tag{20f}
\\
\label{Eq:20g} &\tilde{C}_{mn}^{2}( k ) =C_{mn}^{2}( k ), m,n\in \mathcal{V}, k\ge 0. \tag{20g}
\end{flalign}

Combining Cases I and II, it can be derived that $\tilde{\mathcal{I}}_{h}=\mathcal{I}_{h}$ under the initial value $\tilde{x}_{i}^{0}=x_{i}^{0}+\delta \in \mathcal{S}_{0}^{i}$. Then
\begin{flalign}
\nonumber \mathbf{D}( \mathcal{S}_{0}^{i} ) \ge \underset{\delta \in \mathbb{R}}{\text{sup}}| x_{i}^{0}-\tilde{x}_{i}^{0} |=\underset{\delta \in \mathbb{R}}{\text{sup}}| \delta |=\infty
\end{flalign}
Therefore, the initial value $x_{i}^{0}$ of agent $i$ is preserved against agents $\mathcal{H}$ if agent $i$ has at least one legitimate neighbor $l\in \mathcal{V}\setminus \mathcal{H}$.
\end{proof}
\end{theorem}

\begin{remark}
By \eqref{Eq:19e} and \eqref{Eq:20e}, one knows that the privacy of Algorithm 2 does not have any requirement for the weights $\tilde{C}_{ii}^{1}( 0 )$ and $\tilde{C}_{ll}^{1}( 0 )$. The reason for this is that each agent $i$ in Algorithm 2 does not use such weights in the iterations $k=0,1,\cdots,K$. One benefit of this operation is that it allows the mixing weights of the transmitted information to be used in the iterations $k=0,1,\cdots,K$ without requiring the satisfaction of the sum-one condition, which in turn provides better flexibility in the setting of mixing weights.
\end{remark}

\begin{corollary}
During the running of Algorithm 2, the initial value $x_{i}^{0}$ of agent $i\notin \mathcal{H}$ would be revealed if $\mathcal{N}_{i}^{\text{out}}\cup \mathcal{N}_{i}^{\text{in}}\subset \mathcal{H}$ holds.
\begin{proof}
Recursively computing the update of $x$-variable for $k \le K$ yields
\begin{flalign}
\label{Eq:21} x_i( K+1 ) -x_i( 0 )=\sum_{t=0}^K\!{\sigma ( t )\! \Big(\! \sum_{n\in \mathcal{N}_{i}^{\text{in}}}\!{C_{in}^{1}( t ) x_n( t )}\!-\!\sum_{m\in \mathcal{N}_{i}^{\text{out}}}\!{C_{mi}^{1}( t ) x_i( t )}\! \Big)}. \tag{21}
\end{flalign}
Then, using the column stochasticities of $\mathbf{C}_1( k )$ for $k \ge K+1$ and $\mathbf{C}_2( k )$ for $k \ge 0$, we have
\begin{flalign}
\nonumber &x_i( k ) =C_{ii}^{1}( k ) x_i( k ) +\sum_{m\in \mathcal{N}_{i}^{\text{out}}}{C_{mi}^{1}( k ) x_i( k )}, k \ge K,
\\
\nonumber &y_i( k ) =C_{ii}^{2}( k ) y_i( k ) +\sum_{m\in \mathcal{N}_{i}^{\text{out}}}{C_{mi}^{2}( k ) y_i( k )}, k \ge 0.
\end{flalign}
Combining the above relations with \eqref{Eq:8} and \eqref{Eq:9}, one arrives
\begin{flalign}
\label{Eq:22} &x_i( k ) -x_i( K+1 )=\sum_{t=K+1}^{k-1}\!{\Big( \!\sum_{n\in \mathcal{N}_{i}^{\text{in}}}{C_{in}^{1}( t ) x_n( t )}\!-\!\sum_{m\in \mathcal{N}_{i}^{\text{out}}}{C_{mi}^{1}( t ) x_i( t )} \!\Big)}, \tag{22}
\\
\label{Eq:23} &y_i( k ) -y_i( 0 )=\sum_{t=0}^{k-1}{\Big( \sum_{n\in \mathcal{N}_{i}^{\text{in}}}{C_{in}^{2}( t ) y_n( t )}\!-\!\sum_{m\in \mathcal{N}_{i}^{\text{out}}}{C_{mi}^{2}( t ) y_i( t )} \Big)}. \tag{23}
\end{flalign}
Further, combining the results in \eqref{Eq:21} and \eqref{Eq:22} gives
\begin{flalign}
\nonumber x_i( k ) -x_i( 0 )=&\sum_{t=K+1}^{k-1}{\Big( \sum_{n\in \mathcal{N}_{i}^{\text{in}}}{C_{in}^{1}( t ) x_n( t )}-\sum_{m\in \mathcal{N}_{i}^{\text{out}}}{C_{mi}^{1}( t ) x_i( t )} \Big)}
\\
\label{Eq:24}\,\, &+\sum_{t=0}^K{\sigma ( t ) \Big( \sum_{n\in \mathcal{N}_{i}^{\text{in}}}{C_{in}^{1}( t ) x_n( t )}-\sum_{m\in \mathcal{N}_{i}^{\text{out}}}{C_{mi}^{1}( t ) x_i( t )} \Big)}. \tag{24}
\end{flalign}
Note that each agent $j\in \mathcal{H}$ has access to $\mathcal{I}_{h}$. If $\mathcal{N}_{i}^{\text{out}}\cup \mathcal{N}_{i}^{\text{in}}\subset \mathcal{H}$ holds for legitimate agent $i$, all the information involved on the right sides of \eqref{Eq:23} and \eqref{Eq:24} is accessible to the honest-but-curious agents. Then, using $y_i( 0 )=1$ and \eqref{Eq:23}, agent $j$ can capture $y_i( k )$ for all $k$. Further, as $C_{ij}^{1}( k ) =C_{ij}^{2}( k )$ for $k\ge K+1$, $x_i( k )$ can be inferred correctly by agent $j$ using
\begin{flalign}
\nonumber x_i( k ) =\frac{C_{ji}^{1}( k ) x_i( k )}{C_{ji}^{2}( k ) y_i( k )}y_i( k ).
\end{flalign}
Making use of \eqref{Eq:24}, the desired initial value $x_i( 0 ) =x_{i}^{0}$ is revealed.
\end{proof}
\end{corollary}

\subsection{Performance Against Eavesdropping Attacks}
For the eavesdropping attacks, we make the following assumption.
\begin{assumption}\label{A3}
Consider a strongly connected network $\mathcal{G}$, where an external eavesdropper exists. We assume that the parameter $\sigma(0)$ is not accessible to the eavesdropper.
\end{assumption}

\begin{theorem}
Under Assumptions \ref{A1} and \ref{A3}, the initial values $\{ x_{i}^{0} \} _{i\in \mathcal{V}}$ of all agents can be preserved.
\begin{proof}
From the definition of privacy metric in Section II-D, it is shown that all agents' privacy can be safely protected insofar as $\mathbf{D}( \mathcal{S}_0 ) =\infty$. The available information to the external eavesdropper is given as
\begin{flalign}
\nonumber \mathcal{I}_e=\{ C_{ij}^{1}( k ) x_j( k ) ,C_{ij}^{2}( k ) y_j( k ) | \forall i,j\in \mathcal{V},i\ne j,k\ge 0 \}.
\end{flalign}
The dynamic \eqref{Eq:8} can be reformulated as
\begin{flalign}
\label{Eq:25} \mathbf{x}( k+1 ) =\mathbf{x}( k ) +\sigma ( k ) \mathbf{R}\Delta \mathbf{x}( k ), k \le K, \tag{25}
\end{flalign}
where $\mathbf{R}$ denotes the incidence matrix associated network $\mathcal{G}$, and $\Delta \mathbf{x}( k )$ is a stack vector whose $i$-th element is $C_{mn}^{1}( k ) x_n( k )$ with $(m,n)$ being the $i$-th edge in $\mathcal{E}$. Note that the external eavesdropper is only able to infer all $\{ x_i( 0 ) \} _{i\in \mathcal{V}}$ using $\mathcal{I}_e$. To prove $\mathbf{D}( \mathcal{S}_0 ) =\infty$, it is required to indicate that any initial value $\mathbf{\tilde{x}}( 0 ) \triangleq \mathbf{x}( 0 ) +\Delta \sigma ( 0 ) \mathbf{R}\Delta \mathbf{x}( 0 ) \in \mathcal{S}_0$ makes the information $\tilde{\mathcal{I}}_e$ accessed by the external eavesdropper unchanged, i.e., $\tilde{\mathcal{I}}_e=\mathcal{I}_e$, where $\Delta \sigma ( 0 )$ is any value in $\mathbb{R}$. Hence, we only need to prove that it holds $\tilde{\mathcal{I}}_e=\mathcal{I}_e$ under two different initial states $\mathbf{\tilde{x}}( 0 )$ and $\mathbf{x}( 0 )$. Specifically, one derives $\tilde{\mathcal{I}}_e=\mathcal{I}_e$ if the weights are set as
\begin{flalign}
\label{Eq:26a} &\tilde{C}_{mn}^{1}( 0 ) =C_{mn}^{1}( 0 ) x_n^{0} /\tilde{x}_n^{0}, m,n\in \mathcal{V}, m\ne n, \tag{26a}
\\
\label{Eq:26b} &\tilde{C}_{nn}^{1}( 0 ) \in \mathbb{R}, n\in \mathcal{V},\tag{26b}
\\
\label{Eq:26c} &\tilde{\sigma}( 0 ) = \sigma ( 0 ) + \Delta \sigma ( 0 ), \tag{26c}
\\
\label{Eq:26d} &\tilde{C}_{mn}^{1}( k ) =C_{mn}^{1}( k ), m,n\in \mathcal{V}, k\ge 1, \tag{26d}
\\
\label{Eq:26e} &\tilde{C}_{mn}^{2}( k ) =C_{mn}^{2}( k ), k \ge 0, \tag{26e}
\\
\label{Eq:26f} &\tilde{\sigma}( k ) =\sigma ( k ), k\ge 1. \tag{26f}
\end{flalign}
Further, owing to the fact that the rank of $\mathbf{R}$ is $N-1$ and the nullity of $\mathbf{R}$ is $| \mathcal{E} |-N-1$, one concludes that $\Delta \mathbf{x}( 0 )$ is any vector in $\mathbb{R}^{| \mathcal{E} |}$. In other words, the probability of $\Delta \mathbf{x}( 0 )$ landing in the null space of $\mathbf{R}$ is zero. Thus, for any $n\in \mathcal{V}$, it holds
\begin{flalign}
\nonumber [ \mathbf{R}\Delta \mathbf{x}( 0 ) ] _n =\sum_{m\in \mathcal{N}_{n}^{\text{in}}}{C_{nm}^{1}( 0 ) x_m( 0 )}-\sum_{m\in \mathcal{N}_{n}^{\text{out}}}{C_{mn}^{1}( 0 ) x_n( 0 )}\ne 0.
\end{flalign}
Naturally, $\tilde{x}_n( 0 ) -x_n( 0 ) =[ \Delta \sigma( 0 ) \mathbf{R}\Delta \mathbf{x}( 0 ) ] _n$ can be any value in $\mathbf{R}$. Therefore,
\begin{flalign}
\nonumber \mathbf{D}( \mathcal{S}_0 ) =\underset{\mathbf{x}( 0 ) ,\mathbf{\tilde{x}}( 0 ) \in \mathcal{S}_0}{\text{sup}}\lVert \mathbf{x}( 0 ) -\mathbf{\tilde{x}}( 0 ) \rVert
=\underset{\Delta \sigma ( 0 ) \in \mathbb{R}}{\text{sup}}\lVert \Delta \sigma ( 0 ) \mathbf{R}\Delta \mathbf{x}( 0 ) \rVert =\infty.
\end{flalign}
That is to say, all initial values $\{ x_i( 0 ) \} _{i\in \mathcal{V}}$ are preserved against the external eavesdropper.
\end{proof}
\end{theorem}

\begin{corollary}
If the update rule for $k\le K$ in \eqref{Eq:8} is substituted with \eqref{Eq:7}, Algorithm \ref{alg:2} cannot preserve the initial value of each agent $i$ against eavesdropping attacks.
\begin{proof}
Recursively computing the update of $x$-variable in \eqref{Eq:7} for $k \le K$ gives
\begin{flalign}
\label{Eq:27} x_i( K+1 ) -x_i( 0 )=\sum_{t=0}^K{\Big( \sum_{n\in \mathcal{N}_{i}^{\text{in}}}{C_{in}^{1}( t ) x_n( t )}-\sum_{m\in \mathcal{N}_{i}^{\text{out}}}{C_{mi}^{1}( t ) x_i( t )} \Big)}. \tag{27}
\end{flalign}
Note that \eqref{Eq:22} and \eqref{Eq:23} still hold in this setting. Combining \eqref{Eq:27} with \eqref{Eq:22}, we have
\begin{flalign}
\label{Eq:28} &x_i( k ) -x_i( 0 )=\sum_{t=0}^{k-1}{\Big( \sum_{n\in \mathcal{N}_{i}^{\text{in}}}{C_{in}^{1}( t ) x_n( t )}-\sum_{m\in \mathcal{N}_{i}^{\text{out}}}{C_{mi}^{1}( t ) x_i( t )} \Big)}. \tag{28}
\end{flalign}
Since the external eavesdropper can capture all transmitted information, all terms in the right sides of \eqref{Eq:23} and \eqref{Eq:28} can be accessed by the external eavesdropper. Then, using $y_i( 0 )=1$ and \eqref{Eq:23}, agent $j$ can capture $y_i( k )$ for all $k$. Further, since $C_{ij}^{1}( k ) =C_{ij}^{2}( k )$ for $k\ge K+1$, $x_i( k )$ can be inferred correctly by agent $j$ using
\begin{flalign}
\nonumber x_i( k ) =\frac{C_{ji}^{1}( k ) x_i( k )}{C_{ji}^{2}( k ) y_i( k )}y_i( k ).
\end{flalign}
Making use of \eqref{Eq:28}, the desired initial value $x_i( 0 ) =x_{i}^{0}$ is inferred.
\end{proof}
\end{corollary}

\begin{remark}
According to the discussions above, it is evident that the first $K$-step perturbations are crucial for preserving privacy against honest-but-curious attacks, while the time-varying parameter $\sigma ( t )$ is pivotal in protecting privacy from eavesdropping attacks. Note that we only require that $\sigma ( 0 )$ is agnostic to the eavesdropper. Although this requirement is extremely stringent in practice, it still has some developmental significance. Specifically, due to the arbitrariness of $\sigma ( 0 )$, we can mask only $\sigma ( 0 )$ by some privacy-preserving techniques such as encryption and obfuscation. Since these techniques act only on $\sigma ( 0 )$, they do not impact the convergence of the algorithm.
\end{remark}

\begin{remark}
Theorem 1 states that the randomness of embeddings in the first $K$ iterations has no impact on the consensus performance. Besides, from the privacy analysis, we can see that only changing the mixing weights and auxiliary parameter at the iteration $k=0$ is enough to mask the initial values. That is, we can make Algorithm 2 protect the initial value $x_i( 0 )$ by simply embedding randomness to $\mathbf{C}_1( 0 )$ (i.e., setting $K=1$). Here, our consideration of $K \ge 1$ is to preserve more intermediate states $x_i( k )$, but this also delays the consensus process, see Fig. 3. Therefore, if the intermediate states are not information of privacy concern, we directly set $K = 1$ to obtain the best convergence performance.
\end{remark}

\textbf{Discussion:} The update rules of Algorithm 2 can also be extended to the case of vector states. Actually, privacy (i.e., the agent's initial vector state) is naturally protected provided that each element of the vector state is assigned an independent mixing weights. The details are summarized in Algorithm 3.
\begin{algorithm}[!htb]
    \renewcommand{\thealgorithm}{3}
	\caption{Secure average consensus algorithm in the vector-state case}
	\label{alg:3}
	\begin{algorithmic}[1]
		\STATE \textbf{Input:} Initial states $\mathbf{x}_i( 0 ) =\mathbf{x}_{i}^{0}\in \mathbb{R}^d$ and $y_i( 0 ) =1$ for $i\in \mathcal{V}$; Parameters $K\in \mathbb{N}$, $\mathbf{\Lambda }( k )  \in \mathbb{R}^{d\times d}$ for $k\in \mathbb{N}$, and $\eta \in ( 0,1 )$; Communication network $\mathcal{G}$.
        \STATE \textbf{Weight generation:} See Table 1.
		\FOR{$k=0,1,\cdots$}
        \FOR{$i=1,\cdots,n$ in parallel}
        \STATE Agent $i$ sends the computed $\mathbf{C}_{li}^{1}( k ) \mathbf{x}_i( k )$ and $C_{li}^{2}( k ) y_i( k )$ to $l\in \mathcal{N}_{i}^{\text{out}}$.
        \STATE Agent $i$ uses $\mathbf{C}_{ij}^{1}( k ) \mathbf{x}_j( k )$ and $C_{ij}^{2}( k ) y_j( k )$ from $j\in \mathcal{N}_{i}^{\text{in}}$ to update $\mathbf{x}_i$ and $y_i$ as follows:
        \begin{flalign}
        \label{Eq:29} &\mathbf{x}_i( k+1 ) =\begin{cases}
	\mathbf{x}_i( k ) +\mathbf{\Lambda }( k ) \boldsymbol{\varXi }_i( k ) ,&		\text{if}\,\,k\le K;\\
	\underset{j\in \mathcal{N}_{i}^{\text{in}}\cup \{ i \}}{\sum}{\mathbf{C}_{ij}^{1}( k ) \mathbf{x}_j( k )}&		\text{if}\,\,k\ge K+1.\\
\end{cases} \tag{29}
\\
\label{Eq:30} &y_i( k+1 ) =\underset{j\in \mathcal{N}_{i}^{\text{in}}\cup \{ i \}}{\sum}{C_{ij}^{2}( k ) y_j( k )}, k\ge 0, \tag{30}
        \end{flalign}
        where $\boldsymbol{\varXi }_i( k )\! \triangleq\! \underset{j\in \mathcal{N}_{i}^{\text{in}}}{\sum}\!{\mathbf{C}_{ij}^{1}( k ) \mathbf{x}_j( k )}\!-\!\underset{j\in \mathcal{N}_{i}^{\text{out}}}{\sum}\!{\mathbf{C}_{ji}^{1}( k ) \mathbf{x}_i( k )}$.
        \STATE Agent $i$ computes $\mathbf{z}_i( k+1 ) =\mathbf{x}_i( k+1 ) /y_i( k+1 )$.
        \STATE Until a stopping criteria is satisfied, e.g., agent $i$ stops if $\lVert \mathbf{z}( k ) -\mathbf{1}\otimes \mathbf{\bar{x}}^0 \rVert<\epsilon$ for some predefined $\epsilon >0$, where $\mathbf{\bar{x}}^0= \sum\nolimits_{j=1}^N{\mathbf{x}_j( 0 )}/N$.
        \ENDFOR
        \ENDFOR
    \end{algorithmic}
\end{algorithm}

\begin{table*}[htb]
\label{Tab:1}
\centering
\caption{Parameter design}
\begin{tabular}{|c|c|c|}
  \hline
  \textbf{Parameter}                           & \textbf{Iteration} $k \le K$               & \textbf{Iteration} $k \ge K+1$ \\
  \hline
  $\mathbf{\Lambda }( k )$ & \multicolumn{1}{m{6.2cm}|}{\fontsize{10}{12}\selectfont $\mathbf{\Lambda }( k ) =\text{diag}\{ \sigma _1( k ) ,\cdots ,\sigma _d( k ) \}$, where each $\sigma _l( k )$, $l=1,\cdots d$, is chosen from $\mathbb{R}$ independently} & $\setminus$ \\
  \hline
  $C_{ij}^{2}( k )$        & \multicolumn{2}{m{10cm}|}{\fontsize{10}{12}\selectfont Each $C_{ij}^{2}( k )$ is chosen from $[ \eta ,1 ]$ for $j\in \mathcal{N}_{i}^{\text{in}}\cup \{ i \}$ with satisfying $\sum\nolimits_{i=1}^N{C_{ij}^{2}( k )}=1$}\\
  \hline
  $\mathbf{C}_{ij}^{1}( k )$               & \multicolumn{1}{m{6.2cm}|}{\fontsize{10}{12}\selectfont $\mathbf{C}_{ij}^{1}( k ) =\text{diag}\{ C_{ij,1}^{1}( k ),\cdots ,C_{ij,d}^{1}( k ) \}$, where each $C_{ij,l}^{1}( k )$, $l=1,\cdots d$, is chosen from $\mathbb{R}$ for $i\in \mathcal{N}_{j}^{\text{out}}\cup \{ j \}$ independently}  & \multicolumn{1}{m{3.8cm}|}{\fontsize{10}{12}\selectfont $\mathbf{C}_{ij}^{1}( k ) =C_{ij}^{1}( k ) \mathbf{I}$, where $C_{ij}^{1}( k )$ is equal to $C_{ij}^{2}( k )$} \\
  \hline
\end{tabular}
\end{table*}

\begin{corollary}
 The following statements hold:
\begin{enumerate}[i)]
\item Let $\{ ( \mathbf{z}_i( k ) ) _{i=1}^{N} \} _{k\in \mathbb{N}}$ be the sequence generated by Algorithm 3. Define $\mathbf{\bar{x}}^0=\sum\nolimits_{j=1}^N{\mathbf{x}_{j}^{0}}/N$ as the average initial state. Under Assumption 1, it holds $\lVert \mathbf{z}( k ) -\bar{x}^0\mathbf{1} \rVert \le c_{v}\rho ^k$ for all $k\in \mathbb{N}$, where $c_{v}=\sqrt{d}c$.
\item Let $\mathcal{H}$ denote the set of honest-but-curious agents. Under Assumptions 1-2, the initial value $\mathbf{x}_i( 0 )$ of agent $i\notin \mathcal{H}$ can be preserved against $\mathcal{H}$ during the running of Algorithm 3;
\item Under Assumptions 1 and 3, the initial values $\mathbf{x}_i( 0 )$ of all agents $i\notin \mathcal{H}$ can be preserved against eavesdropping attacks during the running of Algorithm 3.
\end{enumerate}
\begin{proof}
The proof of i) follows a similar path to Theorem 1, the difference lies only in the use of the Kronecker product and thus omitted. Then, by the setup of Table I, it is possible to make each element of the vector state hold an independent coupling weight. According to the analysis of Theorems 2-3, we can know that each scalar-state element in the vector state can be preserved against both honest-but-curious and eavesdropping attacks. Therefore, each vector state can also be preserved.
\end{proof}
\end{corollary}

\section{Experiments Validation}
We construct simulations to confirm the consensus and the privacy performances of our methods. Two directed networks are built in Fig. 2.

\begin{figure}[htbp]
\centering
\subfloat[$\mathcal{G}_1$]{
\label{Fig.2.a}
\includegraphics[trim = 20mm 20mm 20mm 16mm, clip, width=4.5cm]{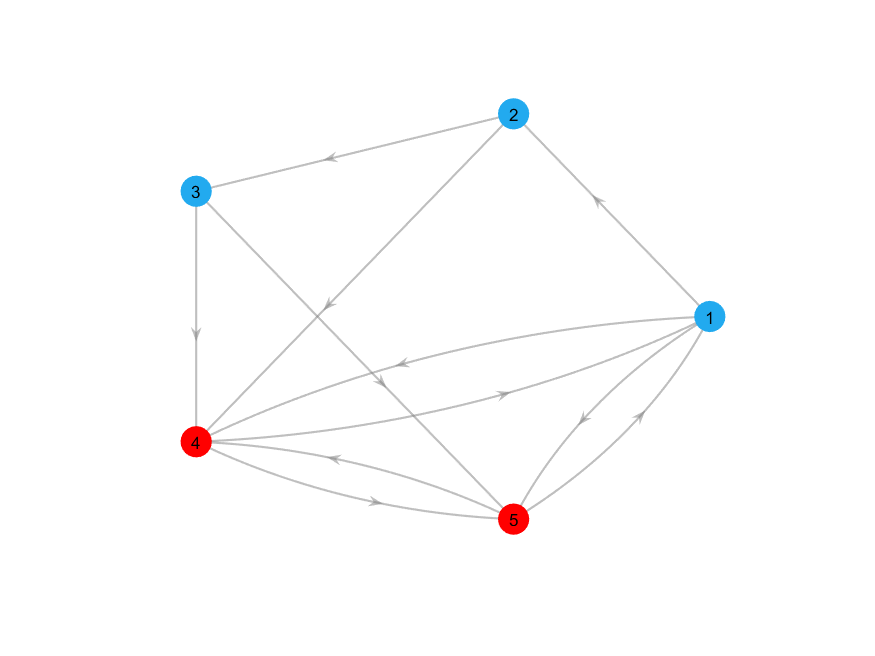}}
\subfloat[$\mathcal{G}_2$]{
\label{Fig.2.b}
\includegraphics[trim = 20mm 20mm 20mm 16mm, clip, width=6cm]{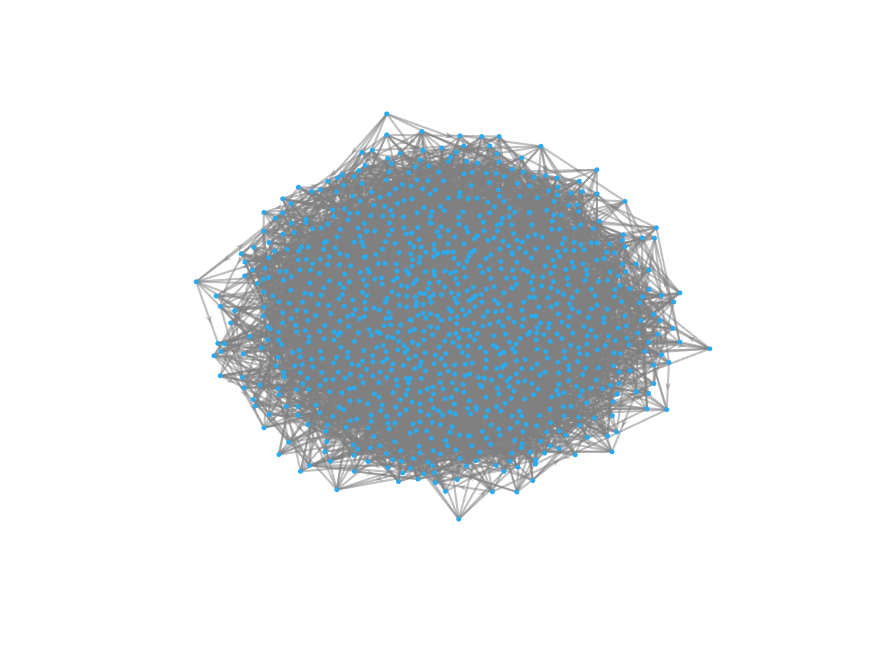}}
\caption{Communication networks. A simple directed network $\mathcal{G}_1$ with $5$ agents and a large-scale directed network $\mathcal{G}_2$ consisting of $1000$ agents.}
\label{fig:2}
\end{figure}


\subsection{Consensus Performance}
\begin{figure}[htbp]
\centering
\subfloat[Scalar case]{
\label{Fig.3.a}
\includegraphics[width=5cm]{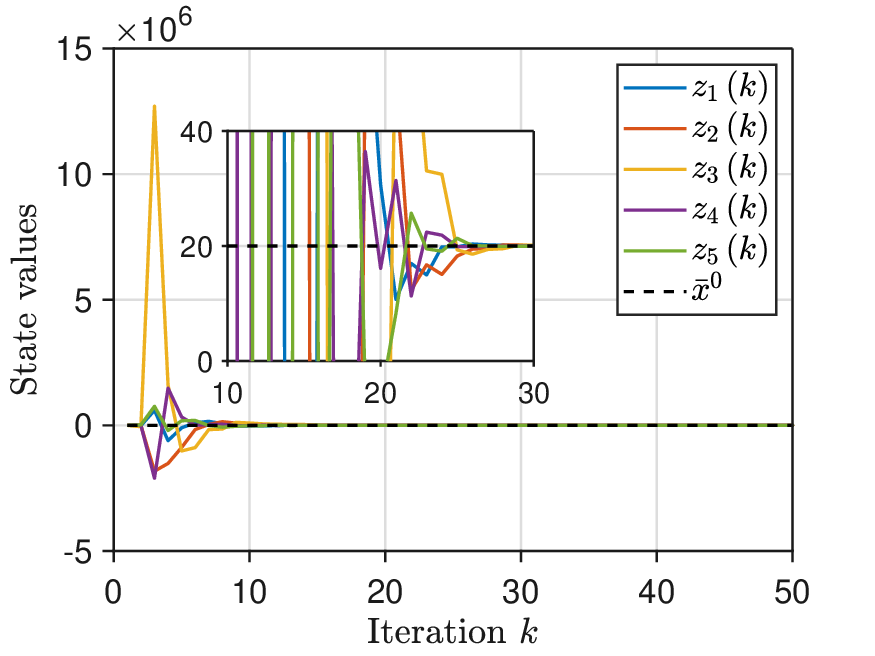}}
\subfloat[Scalar case]{
\label{Fig.3.b}
\includegraphics[width=5cm]{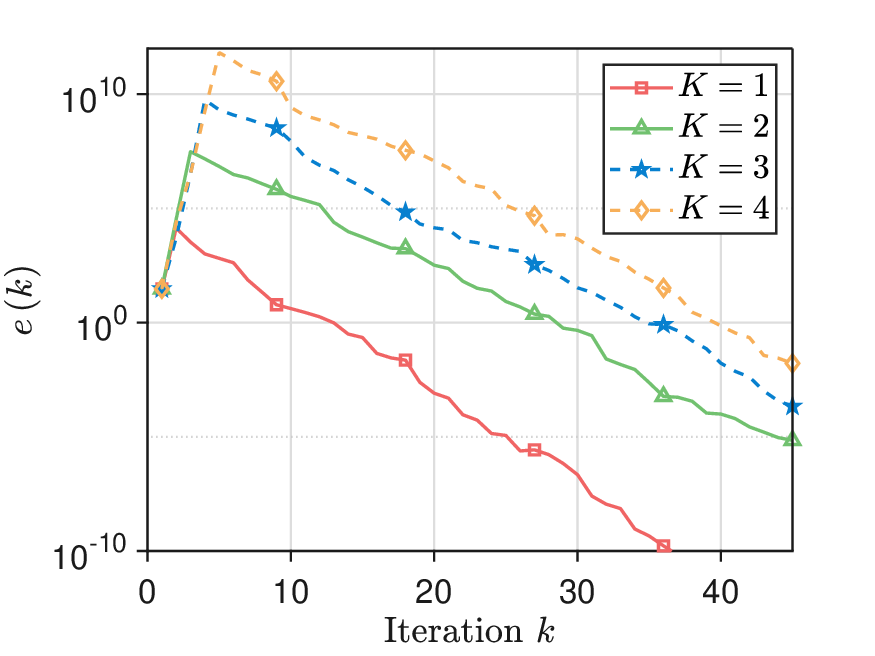}}

\subfloat[Vector case]{
\label{Fig.3.c}
\includegraphics[width=5cm]{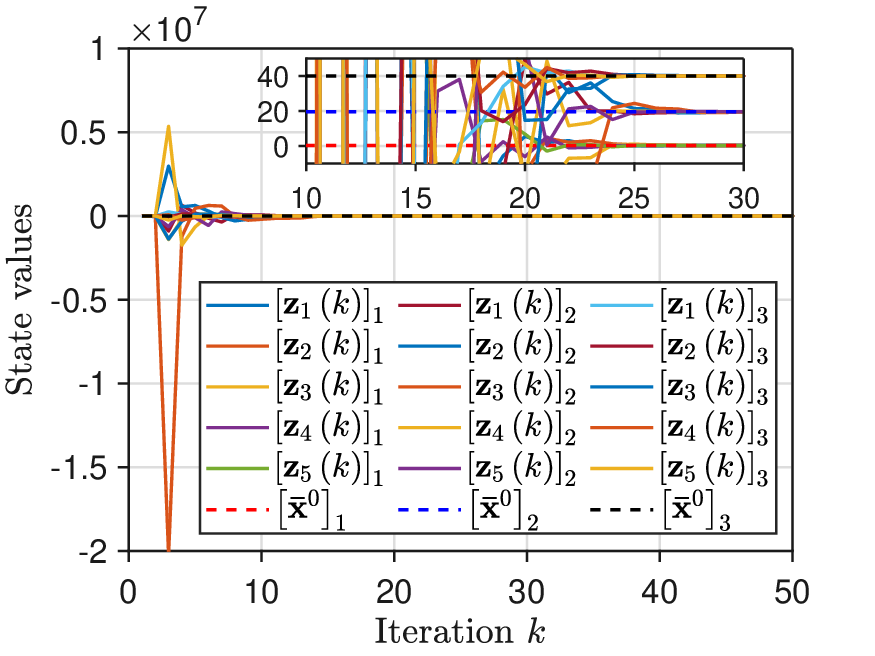}}
\subfloat[Vector case]{
\label{Fig.3.d}
\includegraphics[width=5cm]{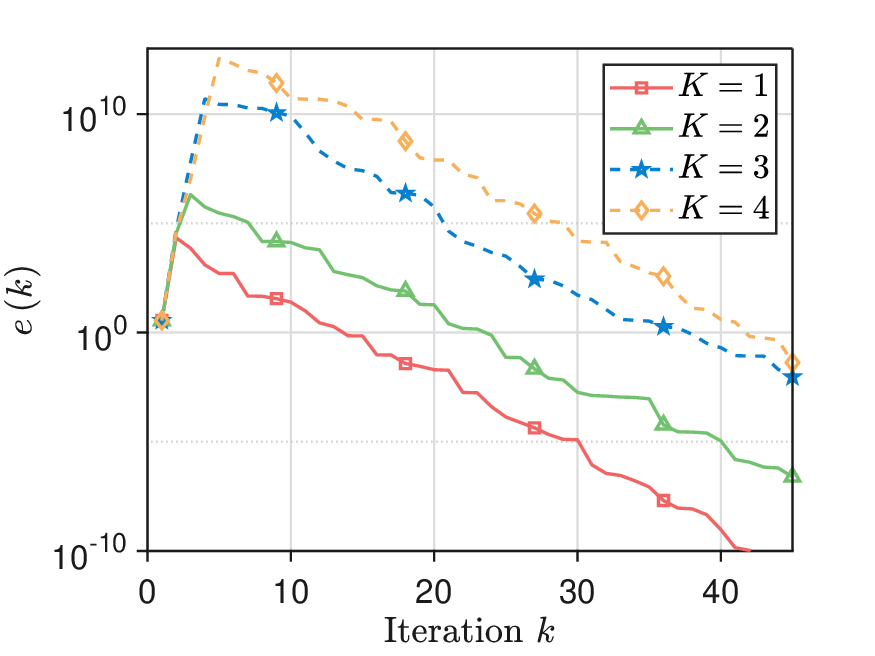}}
\caption{Consensus performance. (a)-(b) The trajectories of states $\{z_i( k )\}$ and the evolutions of $e( k )$ of Algorithm 2; (c)-(d) The trajectories of states $\{\mathbf{z}_i( k )\}$ and the evolutions of $e( k )$ of Algorithm 3.}
\label{fig:3}
\end{figure}

We pick the network $\mathcal{G}_1$ and set $\eta =0.01$. For Algorithm 2, at iteration $k \le K$, the mixing weights $C_{ji}^{1}( k )$ for $j\in \mathcal{N}_{i}^{\text{out}}\cup \{ i \}$ are selected from $( -100,100 )$. The initial states $x_{1}^{0},\cdots ,x_{5}^{0}$ take values of $10,15,20,25,30$, respectively, and thus $\bar{x}^0=20$. The parameter $\sigma ( k )$ is generated from $\mathcal{N}( 0,10 )$ for all $k\le K$. For Algorithm 3, at iteration $k \le K$, the parameters $\sigma _l( k )$ are generated from $\mathcal{N}( 0,10 )$ for $l=1,\cdots ,d$ with $d=3$, and the mixing weights $C_{ij,l}^{1}( k )$ are chosen from $( -100,100 )$ for $l=1,\cdots ,d$ and $j\in \mathcal{N}_{i}^{\text{out}}\cup \{ i \}$. Each component of the initial values $\mathbf{x}_{i}^{0}\in \mathbb{R}^d$, $i=1,\cdots ,5$, is generated from the Gaussian distributions with different mean values $0,20,40$. Fig. \ref{fig:3} plots the evolutionary trajectories of the state variables under $K=2$, and shows the evolutions of $e( k ) =\lVert \mathbf{z}( k ) -\bar{x}^0\mathbf{1} \rVert$ over $K=1,2,3,4$. One observes that: i) Each estimate $z_i( k )$ converges to the average value $\bar{x}^0$, and a linear consensus rate is achieved; and ii) a larger $K$ means a worse consensus accuracy.

\subsection{Comparison with other works}
\begin{figure}[htbp]
\centering
\subfloat[]{
\label{Fig.4.a}
\includegraphics[width=4.5cm]{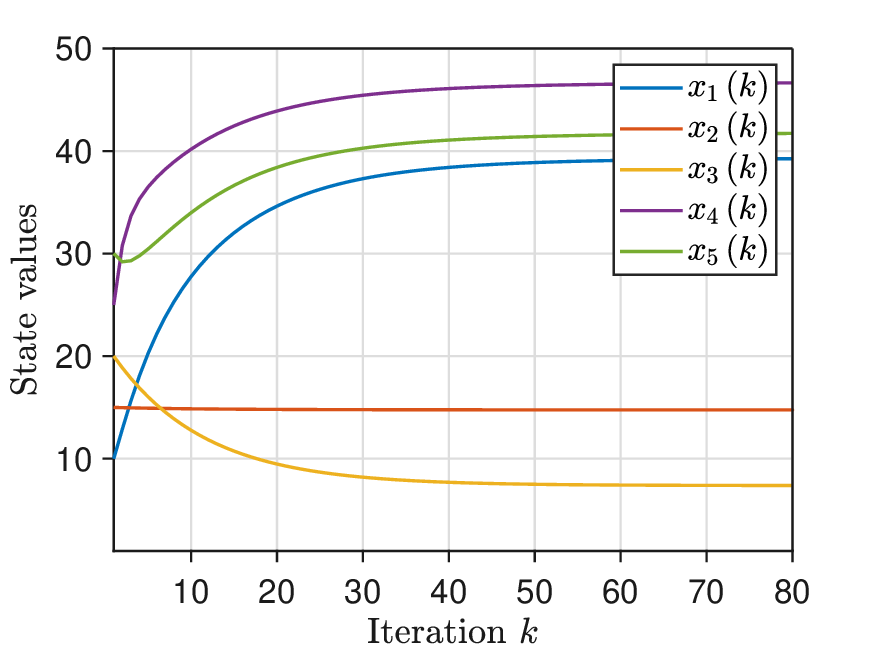}}
\subfloat[]{
\label{Fig.4.b}
\includegraphics[width=4.5cm]{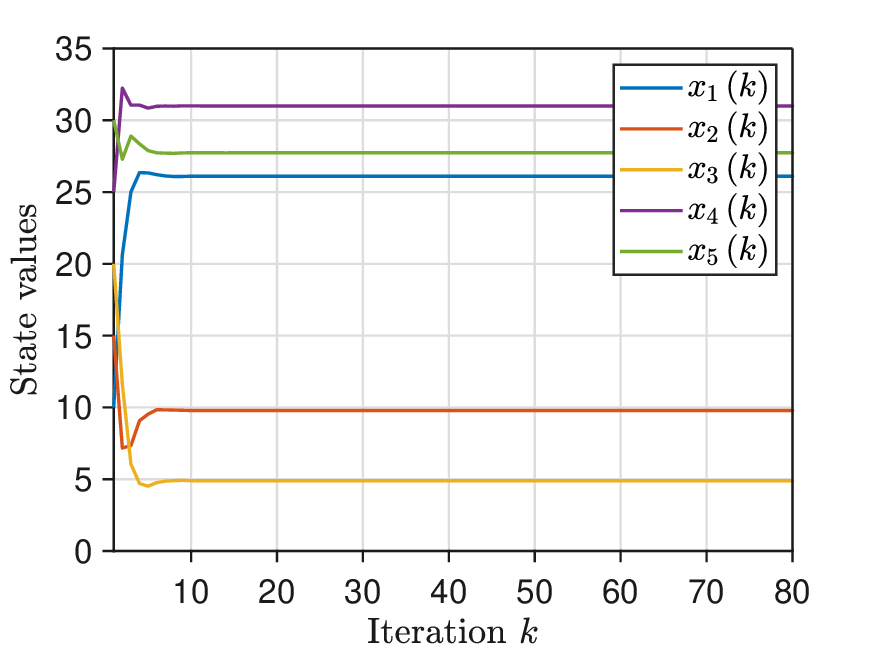}}
\subfloat[]{
\label{Fig.4.b}
\includegraphics[width=4.5cm]{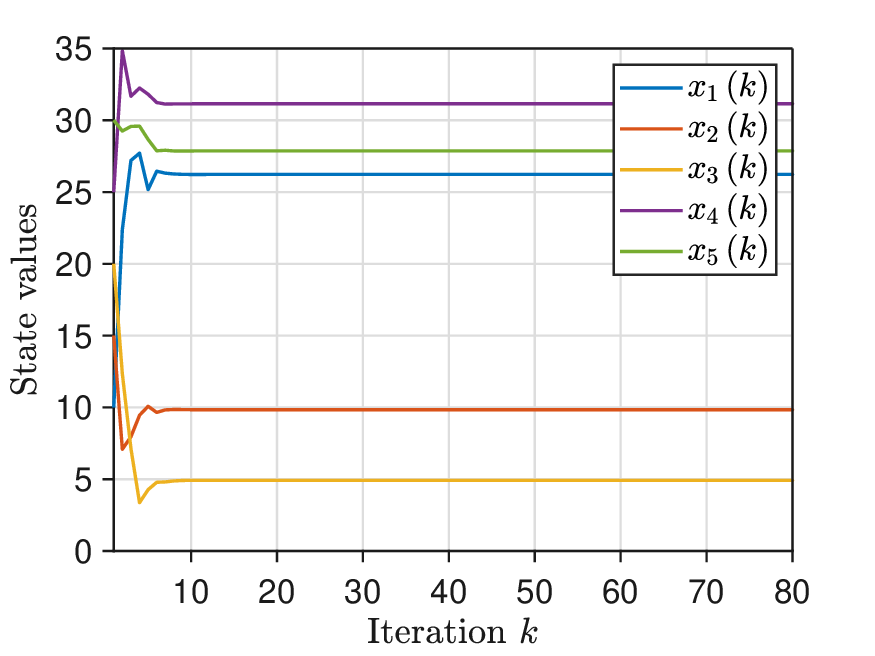}}
\caption{Performance of the other works. (a)-(c) The trajectories of all states $\{x_i( k )\}$ in \cite{Huang2012}, \cite{Mo2017}, \cite{Manitara2013} in order.}
\label{fig:4}
\end{figure}

We compare our algorithms with three data-obfuscation based methods, i.e., the differential privacy algorithm \cite{Huang2012}, the decaying noise algorithm \cite{Mo2017}, and the finite-noise-sequence algorithm \cite{Manitara2013}. Here, we set $K=2$, and the adopted mixing matrix $W$ is generated using the rules in \cite{Huang2012}. Specifically, the element $W_{ij}$ is set to $1/( | \mathcal{N}_{j}^{\text{out}} |+1 )$ if $i\in \mathcal{N}_{j}^{\text{out}}\cup \{ j \} $; otherwise, $W_{ij}=0$. Since the directed and unbalanced networks are more generalizable than the undirected and balanced ones adopted in \cite{Huang2012,Mo2017,Manitara2013}, these algorithms cannot achieve average consensus, as reported in Fig \ref{fig:4}.

\subsection{Effect of network degrees and Scalability}
\begin{figure}[htbp]
\centering
\subfloat[]{
\label{Fig.5.a}
\includegraphics[width=4.5cm]{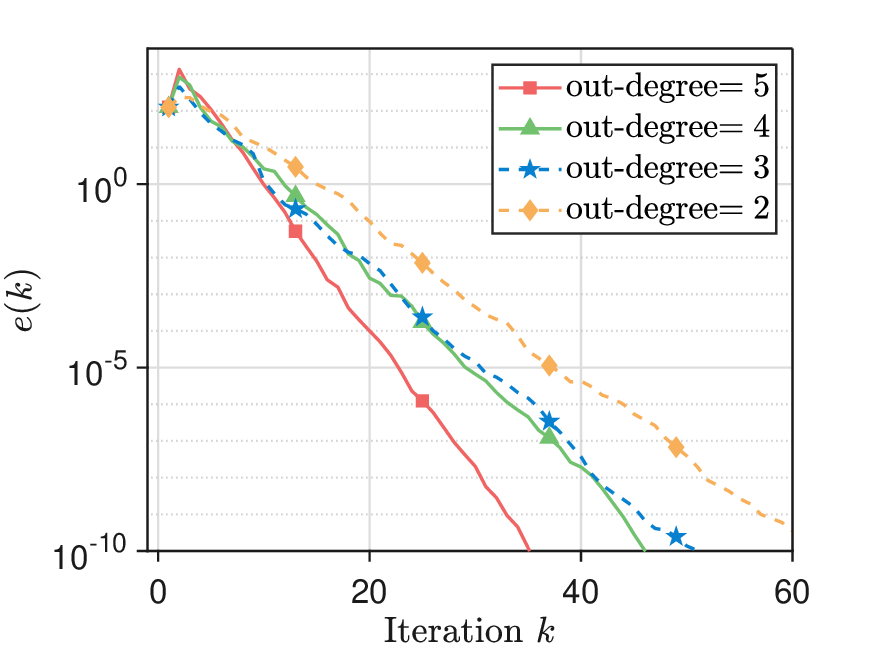}}
\subfloat[]{
\label{Fig.5.b}
\includegraphics[width=4.5cm]{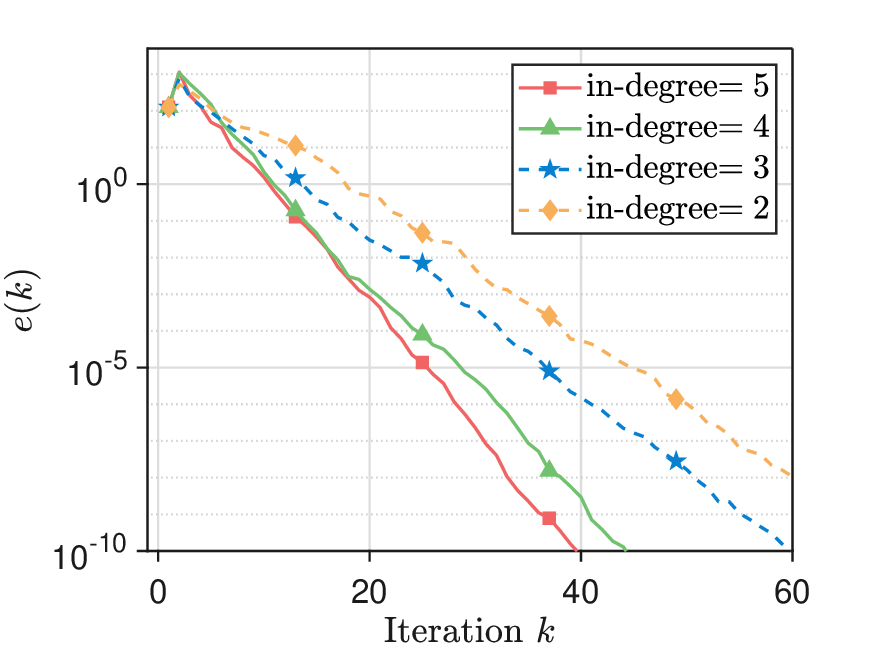}}
\subfloat[]{
\label{Fig.5.b}
\includegraphics[width=4.5cm]{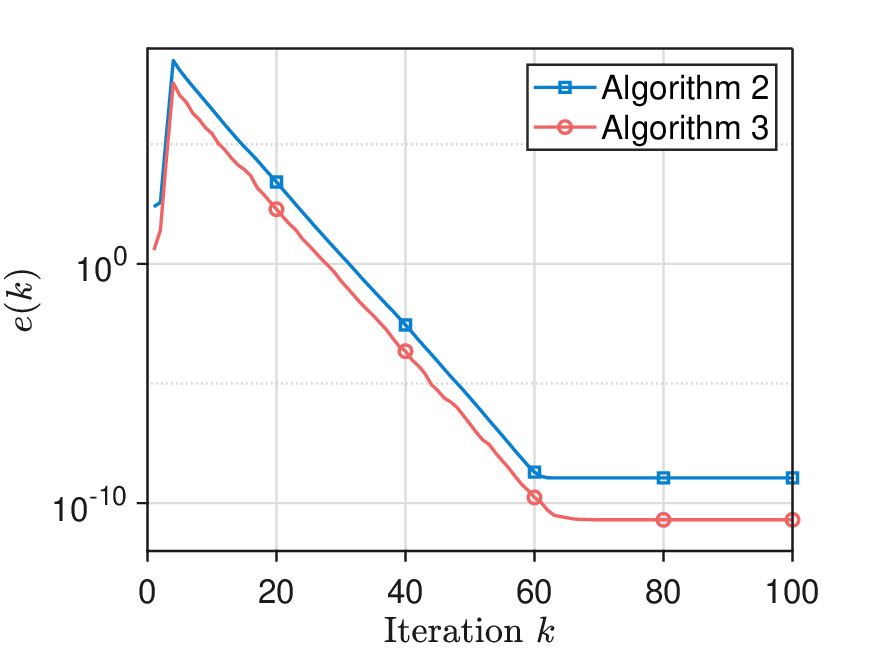}}
\caption{Performance of the proposed algorithm over different works. (a) The effect of out-degrees on consensus rate; (b) The effect of in-degrees on consensus rate; (c) The evolutions of $e( k )$ over the large-scale network $\mathcal{G}_2$.}
\label{fig:5}
\end{figure}

Since the proposed algorithms are performed on unbalanced directed networks, we use different networks to explore the effect of different degrees on the consensus rate. We simulate $8$ communication networks with $10$ agents. Specifically, a directed ring network connecting all agents is built. Then, in the first $4$ networks, each agent arbitrarily selects $D_{i}^{\text{out}}-1=1,2,3,4$ out-neighbors in order; in the last $4$ networks, each agent arbitrarily select $D_{i}^{\text{in}}-1=1,2,3,4$ in-neighbors in order. In this experiment, the variable $x_{1}^{0},\cdots ,x_{10}^{0}$ are taken sequentially from the interval $[10,55]$ at intervals of $5$. We set $K=1$ and $\eta=0.01$. For the iterations $k \le K$, the mixing weights $C_{ji}^{1}( k )$ for $j\in \mathcal{N}_{i}^{\text{out}}\cup \{ i \}$ are selected from $(-5, 5)$, and the parameter $\sigma ( k )$ is generated from $\mathcal{N}( 0,5 )$. Moreover, we employ the network $\mathcal{G}_2$ to demonstrate the scalability of the proposed algorithms. Each initial value $x_{i}^{0}$ or $\mathbf{x}_{i}^{0}$ is generated from i.i.d $\mathcal{N}( 0,1 )$. The parameters $\eta$, $K$, and $d$ take values of $0.05$, $3$, and $10$, respectively. The mixing weights and the parameter $\sigma ( k )$ or $\mathbf{\Lambda }( k )$ are generated in the same way as Section VI-A. As shown in Fig. \ref{fig:5}, it is stated that i) As the out-degree or in-degree increases, Algorithm 2 has a faster consensus rate. A possible reason for this is that the increase in out-degree or in-degree leads to more frequent communication between agents, and thus more information is available for state updates, which in turn leads to a faster consensus rate; ii) The proposed algorithms still ensure that all agents linearly converge to the correct average value even if a large-scale network is used.

\subsection{Privacy Performance}
We evaluate the privacy-preserving performances of Algorithms 2 and 3. Under the network $\mathcal{G}_1$, we consider the initial value of the legitimate agent $1$ will suffer from the joint inference of honest-but-curious agents $4,5$, and agent $2$ is legitimate. In the scalar-state case, we set $x_{1}^{0}=40$ and $x_{2}^{0},\cdots ,x_{N}^{0}$ are generated from the Gaussian distributions with $50$ variance and zero mean, while the initial value $\mathbf{x}_{1}^{0}=[50,50]$ and $\mathbf{x}_{2}^{0},\cdots, \mathbf{x}_{N}^{0}$ are randomly generated from i.i.d. $\mathcal{N}( 0,50 )$ in the vector-state case. Moreover, we set $k=2$ and the maximal iteration $M=200$.

To infer $x_{1}^{0}$, agents $\mathcal{H}=\{ 4,5 \}$ construct some linear equations based on their available information $\mathcal{I}_h=\{ \mathcal{I}_4,\mathcal{I}_5 \}$ outlined below:
\begin{flalign}
\label{Eq:31a} &x_1( k+1 ) -x_1( k ) +\sigma ( k ) C_{21}^{1}( k ) x_1( k )=\sigma ( k ) \Delta x( k ), 0\le k\le K, \tag{31a}
\\
\label{Eq:31b} &x_1( k\!+\!1 ) \!-\!x_1( k ) \!+\!C_{21}^{1}( k ) x_1( k ) \!=\!\Delta x( k ), K\!+\!1\!\le \!k\!\le \!M, \tag{31b}
\\
\label{Eq:31c} &y_1\!( k\!+\!1 ) \!-\!y_1\!( k ) \!+\!C_{21}^{2}( k ) y_1( k ) \!=\!\Delta y( k )\!, 0\!\le \!k\!\le\! M, \tag{31c}
\end{flalign}
where
\begin{flalign}
\nonumber &\Delta x( k ) =\sum_{m\in \{ 4,5 \}}{C_{1m}^{1}( k ) x_m( k )}-\sum_{n\in \{ 4,5 \}}{C_{n1}^{1}( k ) x_1( k )},
\end{flalign}
\begin{flalign}
\nonumber &\Delta y( k ) =\sum_{m\in \{ 4,5 \}}{C_{1m}^{2}( k ) y_m( k )}-\sum_{n\in \{ 4,5 \}}{C_{n1}^{2}( k ) y_1( k )}.
\end{flalign}
Furthermore, agents $\mathcal{H}$ can also construct, for $k=K+1,K+2,\cdots ,M$,
\begin{flalign}
\label{Eq:31d} x_1( k ) -z_1( k ) y_1( k ) =0, \tag{31d}
\end{flalign}
where $z_1( k )$ can be derived from
\begin{flalign}
\nonumber z_1( k ) =\frac{C_{41}^{1}( k ) x_1( k )}{C_{41}^{2}( k ) y_1( k )},
\end{flalign}
since $C_{41}^{1}( k ) = C_{41}^{2}( k )$ for $k \ge K+1$.

\begin{figure}[htbp]
\centering
\subfloat[Scalar case]{
\label{Fig.6.a}
\includegraphics[width=5cm]{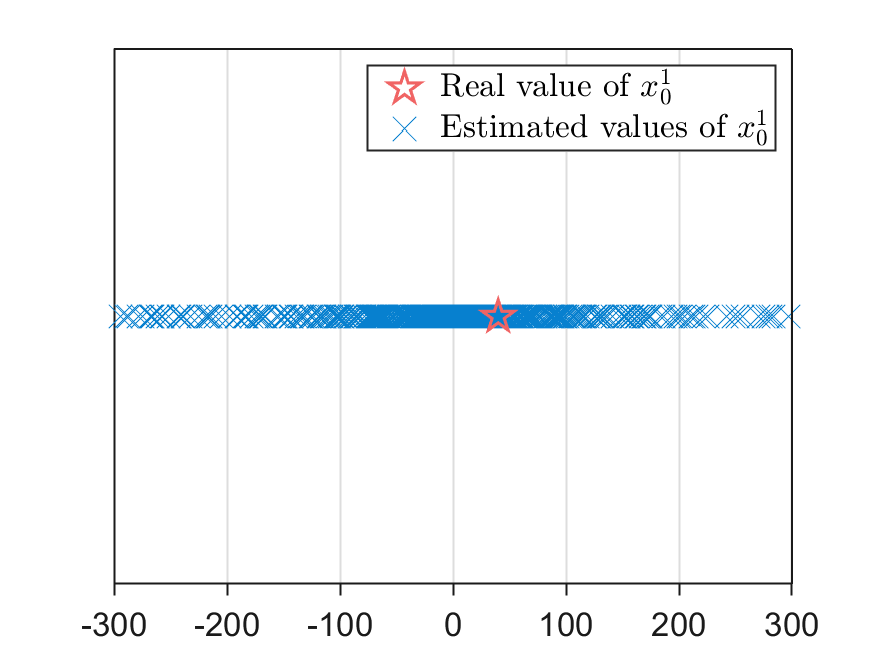}}
\subfloat[Vector case]{
\label{Fig.6.b}
\includegraphics[width=5cm]{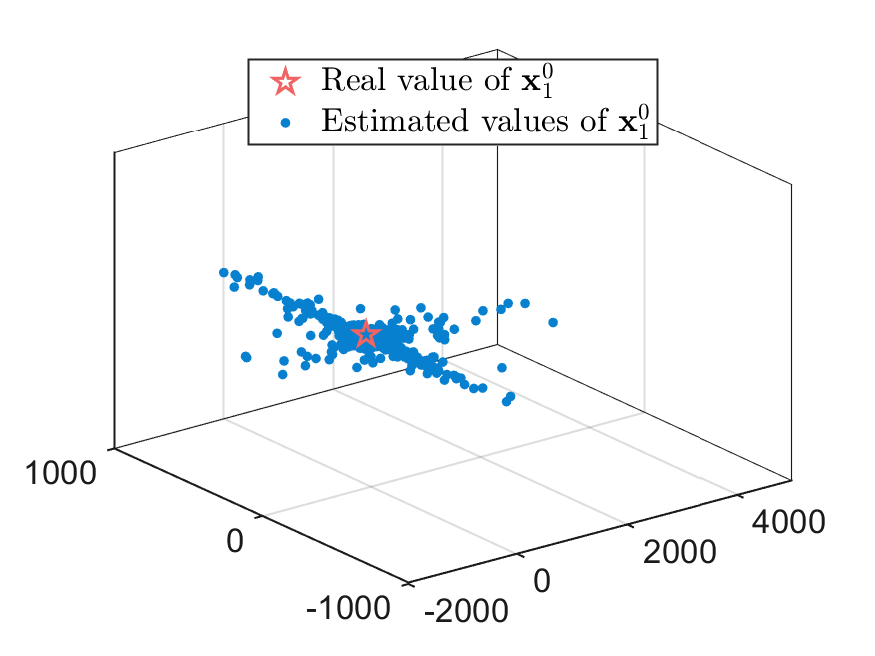}}

\subfloat[Scalar case]{
\label{Fig.6.c}
\includegraphics[width=5cm]{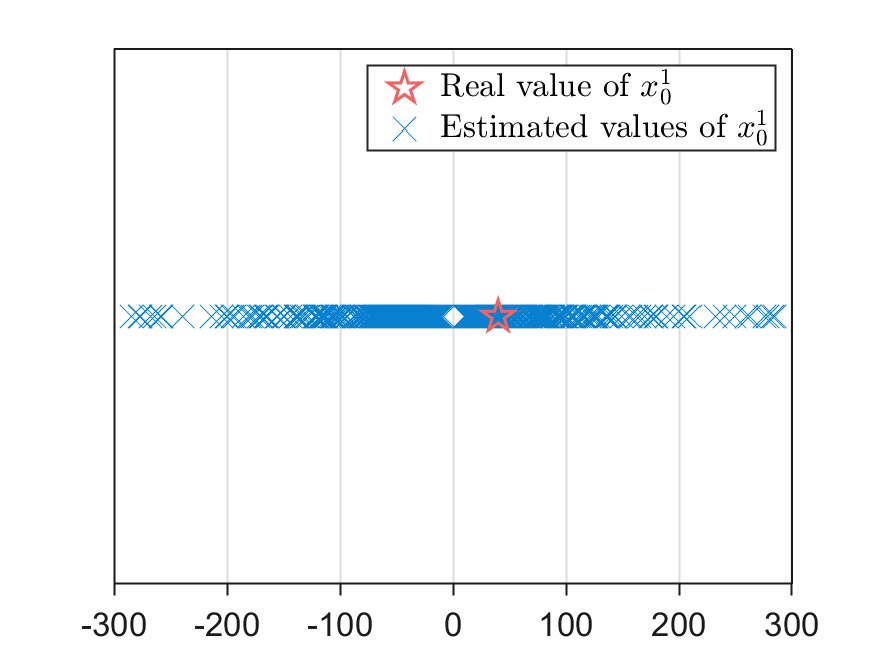}}
\subfloat[Vector case]{
\label{Fig.6.d}
\includegraphics[width=5cm]{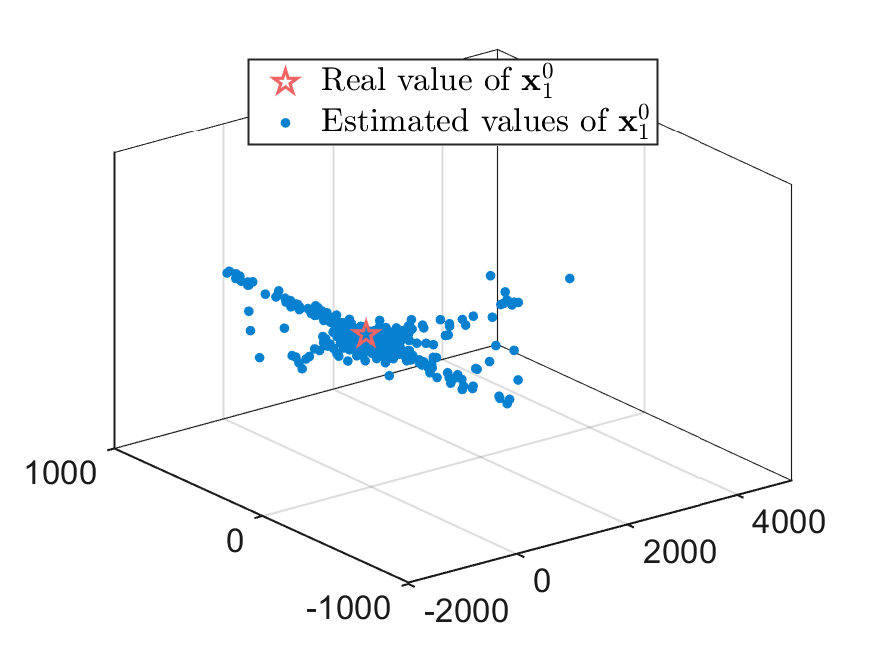}}
\centering
\caption{Privacy Performance. (a) Estimation results of $x_{1}^{0}$ by $\mathcal{H}$; (b) Estimation results of $\mathbf{x}_{1}^{0}$ by $\mathcal{H}$; (c) Estimation results of $x_{1}^{0}$ by the external eavesdropper; (d) Estimation results of $\mathbf{x}_{1}^{0}$ by the external eavesdropper.}
\label{fig:6}
\end{figure}

The number of linear equations is $3M-K+2$ while that of unknown variables to $\mathcal{H}$ is $4M+5$, including $x_1( 0 ), \cdots, x_1( M+1 )$, $C_{21}^{1}( 0 ) x_1( 0 )$, $\cdots$, $C_{21}^{1}( M ) x_1( M )$, $y_1( 1 )$, $\cdots$, $y_1( M+1 )$, and $C_{21}^{2}( 0 ) y_1( 0 ) ,\cdots ,C_{21}^{2}( M ) y_1( M )$. Consequently, there are infinitely many solutions due to the fact that the number of equations is less than that of unknown variables. The analysis of the vector-state case is similar to that of the scalar-state case, so it will not be elaborated here. To uniquely determine $x_{1}^{0}$, we use the least-squares solution to infer $x_{1}^{0}$. In this experiment, agents in $\mathcal{H}$ estimate $x_{1}^{0}$ or $\mathbf{x}_{1}^{0}$ for $1000$ times. Figs. \ref{Fig.6.a}-\ref{Fig.6.b} show the estimated results. One can observe that agents in $\mathcal{H}$ fail to obtain a desired estimate of $x_{1}^{0}$ or $\mathbf{x}_{1}^{0}$.

Next, we consider the case of eavesdropping attacks. The parameter settings follow the above experiment. To infer the value $x_{1}^{0}$, the external eavesdropper constructs some linear equations below based on its available information $\mathcal{I}_e$:
\begin{flalign}
\label{Eq:32a} &x_1( k+1 ) -x_1( k ) =\sigma ( k ) \Delta \hat{x}( k ), 0\le k\le K+1, \tag{32a}
\\
\label{Eq:32b} &x_1( k+1 ) -x_1( k ) =\Delta \hat{x}( k ), K+1\le k\le M, \tag{32b}
\\
\label{Eq:32c} &y_1( k+1 ) -y_1( k ) =\Delta \hat{y}( k ), 0\le k\le M, \tag{32c}
\end{flalign}
where
\begin{flalign}
\nonumber &\Delta \hat{x}( k ) \!=\!\sum_{m\in \{ 4,5 \}}{C_{1m}^{1}( k ) x_m( k )}\!-\!\sum_{n\in \{ 2,4,5 \}}{C_{n1}^{1}( k ) x_1( k )},
\\
\nonumber &\Delta \hat{y}( k ) \!=\!\sum_{m\in \{ 4,5 \}}{C_{1m}^{2}( k ) y_m( k )}\!-\!\sum_{n\in \{ 2,4,5 \}}{C_{n1}^{2}( k ) y_1( k )}.
\end{flalign}
Further, the external eavesdropper can deduce from (32) that
\begin{flalign}
\label{Eq:33a} &x_1( K+1 ) -x_1( 0 ) =\sum_{t=0}^K{\sigma ( t ) \Delta \hat{x}( t )}, \tag{33a}
\\
\label{Eq:33b} &x_1\!( k\!+\!1 ) \!-\!x_1( K\!+\!1 ) \!=\!\sum_{t=K\!+\!1}^k{\Delta \hat{x}( t )}, K\!+\!1\!\le\! k\!\le\! M, \tag{33b}
\\
\label{Eq:33c} &y_1( k+1 ) -y_1( 0 ) =\sum_{t=0}^k{\Delta \hat{y}( t )}, 0\le k\le M. \tag{33c}
\end{flalign}
Obviously, all terms in the right side of (33) can be accessed by the external eavesdropper. Consequently, using $y_1( 0 ) =1$, the eavesdropper can be aware of all $y_1( k )$, $k\in \mathbb{N}$. Moreover, the external eavesdropper can capture $C_{21}^{1}( k ) x_1( k )$ and $C_{21}^{2}( k ) y_1( k )$ for $k=K+1,\cdots ,M$. Then, $x_1( k )$ for $k=K+1,\cdots ,M$ can be derived using
\begin{flalign}
\nonumber x_1( k ) =\frac{C_{21}^{1}( k ) x_1( k )}{C_{21}^{2}( k ) y_1( k )}y_1( k ).
\end{flalign}
This implies that all information in \eqref{Eq:32b} and \eqref{Eq:32c} is captured by the external eavesdropper, which is considerably different from the case of honest-but-curious attacks. So, only \eqref{Eq:32a} has some unknown variables $\sigma ( k )$, $k=0,\cdots ,K$ and $x_1( 0 )$ for the external eavesdropper. The vector-state case leads to the same results as the scalar-state case by following the same analysis path, so it is not stated again. In this experiment, we still use the least-squares solution to estimate $x_{1}^{0}$. The external eavesdropper estimates $x_{1}^{0}$ or $\mathbf{x}_{1}^{0}$ for $1000$ times. Figs. \ref{Fig.6.c}-\ref{Fig.6.d} show the estimated results. One observes that the external eavesdropper cannot obtain an expected estimate of $x_{1}^{0}$ or $\mathbf{x}_{1}^{0}$.

\section{Conclusion}
We proposed a dynamics-based privacy-preserving push-sum algorithm over unbalanced digraphs. We theoretically analyzed its linear convergence rate and proved it can guarantee the privacy of agents against both honest-but-curious and eavesdropping attacks. Finally, numerical experiments further confirmed the soundness of our work.
Future research will consider efforts to prevent eavesdropping attacks under a weaker assumption, as well as consider efforts to protect privacy even after $K$ is removed.

\section*{CRediT authorship contribution statement}
Huqiang Cheng: Methodology, Formal analysis, Writing - original draft. Mengying Xie: Methodology, Writing - review \& editing. Xiaowei Yang: Supervision, Writing - review \& editing. Qingguo L{\"u}: Formal analysis, Writing - review \& editing. Huaqing Li: Writing - review \& editing, Funding acquisition.

\section*{Declaration of competing interest}
The authors declare that they have no known competing financial interests or personal relationships that could have appeared to influence the work reported in this paper.

\section*{Data available}
Data will be made available on request.

\section*{Acknowledgment}
This work is supported by the National Natural Science Foundation of China (62302068 and 61932006).

\newpage
\appendix
\section{Proof of Theorem 1}
\begin{proof}
We divide the convergence analysis into two cases.

\noindent \textbf{Case I:} We consider the case of $k\ge K+2$. It holds $\mathbf{C}_1( k ) =\mathbf{C}_2( k )$. Recalling \eqref{Eq:12} and \eqref{Eq:13}, we have, for $l \ge 1$,
\begin{flalign}
\label{Eq:34} &\mathbf{x}( K+l+1 ) =\mathbf{\Phi }_1( K+l:K+1 ) \mathbf{x}( K+1 ), \tag{34}
\\
\label{Eq:35} &\mathbf{y}( K+l+1 ) =\mathbf{\Phi }_2( K+l:K+1 ) \mathbf{y}( K+1 ). \tag{35}
\end{flalign}
Referring \cite[Corollary 2]{Nedic2015}, there exists a sequence of stochastic vectors $\{ \boldsymbol{\varphi }( k ) \} _{k\in \mathbb{N}}$ such that, for any $i,j\in \mathcal{V}$,
\begin{flalign}
\nonumber | [ \mathbf{\Phi }_1( k:K+1 ) ] _{ij}-\varphi _i( k ) |\le c_0\rho ^{k-K-1},
\end{flalign}
where $c_0=2( 1+\rho ^{-N+1} ) /( 1-\rho ^{N-1} )$ and $\rho =( 1-\eta ^{N-1} ) ^{\frac{1}{N-1}}$. Moreover, $\varphi _i( k ) \ge \eta ^N/N$. Thus, it follows that, for $l\ge 1$,
\begin{flalign}
\label{Eq:36} | [ \mathbf{M}( K+l:K+1 ) ] _{ij} |\le c_0\rho ^{l-1}. \tag{36}
\end{flalign}
where $\mathbf{M}( K+l:K+1 ) \triangleq \mathbf{\Phi }_1( K+l:K+1 ) -\boldsymbol{\varphi }( K+l ) \mathbf{1}^{\top}$. Since $\mathbf{C}_1( k ) =\mathbf{C}_2( k )$, it holds that $\mathbf{\Phi }_1( K+l:K+1 ) =\mathbf{\Phi }_2( K+l:K+1 )$ for $l\ge 1$. So \eqref{Eq:34} and \eqref{Eq:35} can be evolved as
\begin{flalign}
\label{Eq:37} &\mathbf{x}( K+l+1 ) =\mathbf{M}( K+l:K+1 ) \mathbf{x}( K+1 ) +\boldsymbol{\varphi }( K+l ) \mathbf{1}^{\top}\mathbf{x}( K+1 ), \tag{37}
\\
\label{Eq:38} & \mathbf{y}( K+l+1 ) =\mathbf{M}( K+l:K+1 ) \mathbf{y}( K+1 )
+N\boldsymbol{\varphi }( K+l ), \tag{38}
\end{flalign}
It follows from \cite[Corollary 2]{Nedic2015} that $y_i( k+1 ) =[ \mathbf{M}( k:0 ) \mathbf{1} ] _i+N\varphi _i( k ) \ge \eta ^N$ for any $k\in \mathbb{N}$.
Using the relation \eqref{Eq:16}, one arrives
\begin{flalign}
\label{Eq:39} \bar{x}^0=\frac{\sum\nolimits_{j=1}^N{x_j( 0 )}}{N}=\frac{\mathbf{1}^{\top}\mathbf{x}( 0 )}{N}=\frac{\mathbf{1}^{\top}\mathbf{x}( K+1 )}{N}. \tag{39}
\end{flalign}
Combining \eqref{Eq:37} and \eqref{Eq:38} with \eqref{Eq:39} yields
\begin{flalign}
\nonumber &\frac{x_i( K+l+1 )}{y_i( K+l+1 )}-\bar{x}^0
\\
\nonumber \,\, =&\frac{x_i( K+l+1 )}{y_i( K+l+1 )}-\frac{\mathbf{1}^{\top}\mathbf{x}( K+1 )}{N}
\\
\nonumber =&\frac{[ \mathbf{M}( K\!+\!l:K\!+\!1 ) \mathbf{x}( K\!+\!1 ) ] _i\!+\!\varphi _i( k\!+\!l ) \mathbf{1}^{\top}\mathbf{x}( K\!+\!1 )}{y_i( K+l+1 )}
-\frac{Q( K;i )}{Ny_i( K+l+1 )}
\\
\nonumber =&\frac{[ \mathbf{M}( K+l:K+1 ) \mathbf{x}( K+1 ) ] _i}{y_i( K+l+1 )}
-\frac{\mathbf{1}^{\top}\mathbf{x}( K+1 ) [ \mathbf{M}( K+l:K+1 ) \mathbf{y}( K+1 ) ] _i}{Ny_i( K+l+1 )},
\end{flalign}
where
\begin{flalign}
\nonumber Q( K;i ) \triangleq &\mathbf{1}^{\top}\mathbf{x}( K+1 ) [ \mathbf{M}( K+l:K+1 ) \mathbf{y}( K+1 ) ] _i+N\varphi _i( k+l ) \mathbf{1}^{\top}\mathbf{x}( K+1 ).
\end{flalign}
Then, we can bound $| z_i( K+l+1 ) -\bar{x}^0 |$ as
\begin{flalign}
\nonumber\,\,&| z_i( K+l+1 ) -\bar{x}^0 |
\\
\nonumber \le &\frac{| [ \mathbf{M}( K+l:K+1 ) \mathbf{x}( K+1 ) ] _i |}{y_i( K+l+1 )}
\\
\nonumber \,\,&+\frac{| \mathbf{1}^{\top}\mathbf{x}( K+1 ) [ \mathbf{M}( K+l:K+1 ) \mathbf{y}( K+1 ) ] _i |}{Ny_i( K+l+1 )}
\\
\nonumber \le &\frac{1}{\eta ^N}\!\Big( \underset{j}{\max}| [ \mathbf{M}( K\!+\!l:K\!+\!1 ) ] _{ij} | \Big) \!\lVert \mathbf{x}( K\!+\!1 ) \rVert _1\!+\!\frac{1}{N\eta ^N}\times
\\
\nonumber \,\, &| \mathbf{1}^{\top}\mathbf{x}( K\!+\!1 ) |\Big( \underset{j}{\max}| [ \mathbf{M}( K\!+\!l:K\!+\!1 ) ] _{ij} | \Big) \lVert \mathbf{y}( K\!+\!1 ) \rVert _1
\\
\nonumber \le &\frac{2}{\eta ^N}\Big( \underset{j}{\max}| [ \mathbf{M}( K+l:K+1 ) ] _{ij} | \Big) \lVert \mathbf{x}( K+1 ) \rVert _1,
\end{flalign}
where the second inequality uses the relation $y_i( K+l+1 ) \ge \eta ^N$, and the last inequality is based on $\lVert \mathbf{y}( K+1 ) \rVert _1=\sum\nolimits_{i=1}^N{| y_i( K+1 ) |}=\mathbf{1}^{\top}\mathbf{y}( K+1 ) =N$ and $| \mathbf{1}^{\top}\mathbf{x}( K+1 ) |\le \lVert \mathbf{x}( K+1 ) \rVert _1$. Further taking into account \eqref{Eq:36}, one derives that
\begin{flalign}
\nonumber | z_i( K+l+1 ) -\bar{x}^0 |\le 2\eta ^{-N}c_0\lVert \mathbf{x}( K+1 ) \rVert _1\rho ^{l-1}.
\end{flalign}
Thus, we arrive that
\begin{flalign}
\label{Eq:40} \lVert \mathbf{z}( K+l+1 ) -\bar{x}^0\mathbf{1} \rVert \le c_1\rho ^{K+l+1}, \tag{40}
\end{flalign}
where $c_1=2\sqrt{N}c_0\lVert \mathbf{x}( K+1 ) \rVert _1\eta ^{-N}\rho ^{-K-2}$. Consequently, for $k\ge K+2$, we have $\lVert \mathbf{z}( k ) -\bar{x}^0\mathbf{1} \rVert \le c_1\rho ^k$.

\noindent \textbf{Case II:} We consider the case of $k\le K+1$. Using $y_i( k+1 ) =[ \mathbf{M}( k:0 ) \mathbf{1} ] _i+N\varphi _i( k ) \le \eta ^N$, one has
\begin{flalign}
\nonumber\,\,  &\frac{x_i( k )}{y_i( k )}-\bar{x}^k=\frac{x_i( k )}{y_i( k )}-\frac{\mathbf{1}^{\top}\mathbf{x}( k )}{N}
\\
\nonumber =&\frac{x_i( k )}{y_i( k )}-\frac{\mathbf{1}^{\top}\mathbf{x}( k ) ( [ \mathbf{M}( k-1:0 ) \mathbf{1} ] _i+N\varphi _i( k-1 ) )}{Ny_i( k )}.
\end{flalign}
Then, we compute $| z_i( k ) -\bar{x}^k |$ as
\begin{flalign}
\nonumber \,\, &| z_i( k ) -\bar{x}^k |
\\
\nonumber \le &\frac{| x_i( k ) |}{y_i( k )}+\frac{| \mathbf{1}^{\top}\mathbf{x}( k ) [ \mathbf{M}( k-1:0 ) \mathbf{1} ] _i |}{Ny_i( k )} +\frac{| \mathbf{1}^{\top}\mathbf{x}( k ) \varphi _i( k-1 ) |}{y_i( k )}
\\
\nonumber \le &\frac{1}{\eta ^N}| x_i( k ) |+\frac{1}{N\eta ^N}| \mathbf{1}^{\top}\mathbf{x}( k ) |\Big( \underset{j}{\max}| [ \mathbf{M}( k-1:0 ) ] |_{ij} \Big)
\\
\nonumber \,\, &+\frac{1}{\eta ^N}| \mathbf{1}^{\top}\mathbf{x}( k ) |( \underset{i}{\max}\,\,\varphi _i( k-1 ) )
\\
\nonumber \le &\frac{1}{\eta ^N}\lVert \mathbf{x}( k ) \rVert _1+\frac{1}{N\eta ^N}\lVert \mathbf{x}( k ) \rVert _1c_0\rho ^{k-1} +( \frac{1}{\eta ^N}-\frac{( N-1 )}{N} ) \lVert \mathbf{x}( k ) \rVert _1,
\end{flalign}
where the last inequality uses the relation $\varphi _i( k-1 ) \ge \frac{\eta ^N}{N}$ for all $i\in \mathcal{V}$ and $k \ge 1$. Specifically, as $\boldsymbol{\varphi }( k )$ is a stochastic vector, $\sum\nolimits_{i=1}^N{\varphi _i( k )}=1$ holds, which in turn gives $\max _{i\in \mathcal{V}}\,\,\varphi _i( k-1 ) \le 1-( N-1 ) \eta ^N/N$. Thus, it yields that
\begin{flalign}
\nonumber \,\, &\lVert \mathbf{z}( k ) -\bar{x}^k\mathbf{1} \rVert
\\
\nonumber \le &\sqrt{N}\eta ^{-N}\lVert \mathbf{x}( k ) \rVert _1+N^{-1/2}\eta ^{-N}\lVert \mathbf{x}( k ) \rVert _1c_0\rho ^{k-1}+\sqrt{N}\eta ^{-N}\lVert \mathbf{x}( k ) \rVert _1
\\
\nonumber \le &c_2\lVert \mathbf{x}( k ) \rVert _1+c_3\lVert \mathbf{x}( k ) \rVert _1\rho ^k,
\end{flalign}
where $c_2=2\sqrt{N}\eta ^{-N}-( N-1 ) /\sqrt{N}$ and $c_3=N^{-1/2}\eta ^{-N}c_0\rho ^{-1}$.

Combining Cases I and II and defining
\begin{flalign}
\label{Eq:41}\!\!\!\! c\!\triangleq\! \max \!\bigg\{ \!\!\!\!\begin{array}{c}
	c_1,( c_2\!+\!c_3 ) \lVert \mathbf{x}( 0 ) \rVert _1,( c_2\rho ^{-1}\!+\!c_3 ) \lVert \mathbf{x}( 1 ) \rVert _1,\\
	\cdots ,( c_2\rho ^{-K-1}\!+\!c_3 ) \lVert \mathbf{x}( K\!+\!1 ) \rVert _1\\
\end{array} \!\!\!\!\bigg\}, \tag{41}
\end{flalign}
one derives, for all $k\in \mathbb{N}$,
\begin{flalign}
\nonumber \lVert \mathbf{z}( k ) -\bar{x}^0\mathbf{1} \rVert \le c\rho ^k,
\end{flalign}
which is the desired result.
\end{proof}

\end{document}